\documentclass[preprint,amsmath,amssymb,aps,pra]{revtex4-2}

\usepackage[utf8]{inputenc}
\usepackage{graphicx}
\usepackage{float}
\usepackage{dcolumn}
\usepackage{caption}
\usepackage{bm}
\usepackage{amsmath}
\usepackage{braket}

\usepackage{color}
\usepackage{upgreek}

\usepackage{ulem}
\usepackage{url}
\usepackage{hyperref}

\usepackage{cleveref}
\usepackage{setspace}

\newcolumntype{L}[1]{>{\raggedright\let\newline\\\arraybackslash\hspace{0pt}}m{#1}}
\newcolumntype{C}[1]{>{\centering\let\newline\\\arraybackslash\hspace{0pt}}m{#1}}
\newcolumntype{R}[1]{>{\raggedleft\let\newline\\\arraybackslash\hspace{0pt}}m{#1}}

\definecolor{dgreen}{rgb}{0.0, 0.5, 0.0}
\definecolor{purple}{rgb}{0.87, 0.0, 1.0}
\newcommand{\red}[1]{{\color{red} #1}}
\newcommand{\blue}[1]{{\color{blue} #1}}

\newcommand{\mycomment}[1]{}

\newcommand{\threej}[6]{
\begingroup
\footnotesize{
\renewcommand*{\arraystretch}{0.6}
\begin{pmatrix}
   #1 & #2 & #3 \\
   #4 & #5 & #6 
\end{pmatrix}}
\endgroup
}

\begin{document}

\title{Ab initio study on the dynamics and spectroscopy of collective rovibrational polaritons}

\author{Tam\'as Szidarovszky}
    \email{tamas.janos.szidarovszky@ttk.elte.hu}
    \affiliation{%
    Institute of Chemistry,
	ELTE E\"otv\"os Lor\'and University, H-1117 Budapest, P\'azm\'any P\'eter s\'et\'any 1/A, Hungary}%


\begin{abstract}
Accurate rovibrational molecular models are employed to gain insight in high-resolution into the collective effects and intermolecular processes arising when molecules in the gas phase are interacting with a resonant infrared (IR) radiation mode.
An efficient theoretical approach is detailed and numerical results are presented for the HCl, H$_2$O and CH$_4$ molecules confined in an IR cavity.
It is shown that by employing a rotationally resolved model for the molecules, revealing the various cavity-mediated interactions between the field-free molecular eigenstates, it is possible to obtain a detailed understanding of the physical processes governing the energy level structure, absorbtion spectra, and dynamic behavior of the confined systems.
Collective effects, arising due to the cavity-mediated interaction between molecules, are identified in energy level shifts, in intensity borrowing effects in the absorbtion spectra, and in the intermolecular energy transfer occurring during Hermitian or non-Hermitian time propagation.
\end{abstract}

\maketitle

\clearpage

\section{\label{introduction}Introduction}

Polaritonic chemistry deals with processes and phenomena that occur when molecules strongly interact with quantized light fields \cite{Ebbesen_review_2016,Zhou_review_2018,Feist_GarciaVidal_ACSphotonics_2018,Edina1,Herrera_PRL_2016,Hertzog2019,Herrera_2020,Kowalewski2017}.
The strong coupling is usually realized by confining the system into a microscale Fabry-Pérot or plasmonic cavity, where the photonic modes can efficiently couple with either electronic or vibrational molecular states, depending on the cavity mode wavelengths.
When the light-matter coupling is stronger than the loss of the cavity mode and the decay rates in the matter, hybrid light-matter states, called polaritons, are formed.
The multidisciplinary field of polaritonic chemistry, which merges molecular sciences with quantum optics has attracted a great deal of attention from chemists and physicists alike, as the formation of polaritons, coherent hybrid light-matter states, have been shown to have an impact on a vast range of material properties and dynamic processes.
For example, electronic polaritons offer new potential energy surfaces (PES) steering photochemistry \cite{Kowalewski2017,Feist_GarciaVidal_ACSphotonics_2018,18SzHaCsCe_a,Fbri2021,22FrGaFe,Bhuyan2023}, vibrational polaritons can modify thermal chemical reactivity \cite{Hirai_2020_review,Nagarajan2021,Simpkins2021,Wang2021,Sidler2022,Li2022,Dunkelberger2022,Schfer2022,Schfer2022b}, and in general polaritons can facilitate rapid intermolecular energy transfer \cite{cavity_Zhong_AngChem_2016,Xiang_2020,Li2021,Gmez2023,Tibben2023}.
Among the many investigated aspects of molecular polaritons, their spectroscopy has also been the subject of many works, both experiments \cite{Shalabney2015,Fleischer_2019,Xiang_2020,Yang2020,Xiang2019,Xiang2018,Dunkelberger_2016,Grafton_2021, Takele_2021,Xiong2023_review,Simpkins2023,Wright2023,Wright2023_b} and theory \cite{Strashko2016,Saurabh2016,FRibeiro2018,Hernandez_2019,Triana_2020,Ribeiro_2021,Fischer_2021,Yu2022,Bonini2022,21SzBaHaVi,GallegoValencia2024,Fischer2024,Simpkins2023,Horak2024}.

In addition to the spectroscopic experiments of molecules exhibiting vibrational strong coupling in the condensed phase \cite{Dunkelberger_2016, Xiang2018, cavity_Muallem_JPCL_2016, Fleischer_2019, Menghrajani2019, Xiang_2020, Grafton_2021, Xiong2023_review}, gas-phase experiments on molecular rovibrational polaritons were also carried out recently in centimeter-scale Fabry-Pérot cavities \cite{Wright2023,Wright2023_b}.
Transfering molecular polaritonics to the gas phase circumvents the complexity introduced by strong intermolecular interactions, allowing for high-resolution spectroscopy and a more detailed understanding of polaritonic states.
For supporting experiments exhibiting vibrational strong coupling, a variety of theoretical tools are already at hand, such as models on vibrational polaritons that incorporate linear response theory \cite{Bonini2022,YuenZhou2024}, vibrational configuration interaction \cite{Yu2022,Yu2024}, or parametrized single \cite{Hernandez_2019,Triana_2020,Fischer_2021} or multiple \cite{Ribeiro_2021,Saurabh2016,FRibeiro2018} anharmonic vibrational modes, some of which include empirically adjustable parameters to include decay mechanisms.
It is also important to point out that the reported experimental cavity spectra of Refs. \cite{Wright2023,Wright2023_b} can be perfectly reproduced using classical expressions for the transmission of light through a Fabry-Pérot cavity containing a dispersive medium, showing that the quantum electrodynamic treatment of radiation is not always necessary (note that the high-resolution quantized data for the molecule is included in the classical description).
This fact might not come as a complete surprise, considering that the Rabi splitting of molecular levels, induced by classical light, i.e., Autler--Townes splitting \cite{AutlerTownes_original}, has been known and used for a long time to help assign rovibrational spectral peaks \cite{AutlerTownes_in_spectroscopy_1997} or deduce molecular parameters \cite{AutlerTownes_in_spectroscopy_2017,AutlerTownes_in_spectroscopy_2012}.
Also, recent results suggest that starting from a quantum theoretical description of polaritonic systems, under certain approximations, one might reach expressions very similar to the classical expressions for the spectral properties of a ``cavity radiation + molecules" system \cite{24Schwennicke}.
Nonetheless, there are aspects of polaritons (cavity-mediated energy transfer and intra/intermolecular couplings, population transfer between bright- and dark states, etc.), which might not be captured using classical formulae, rendering the development in the quantized description of polaritonic systems also necessary.

For molecules freely rotating in the gas phase %, most existing methods can not be directly used, becuase
the appropriate treatment of molecular rotations in addition to vibrations is also needed.
Not only is this the case because the rotational fine structure in the energetics is important in high resolution, but also because the coupling strength between the transition dipole and the radiation field, having laboratory-fixed polarization, depends on the molecular orientation.
When molecules are coherently coupled by the cavity radiation, simple orientational averaging is in principle erroneous \cite{LICI3,LICI5,18SzHaCsCe_b}, thus explicitly including the rotational degrees of freedom in the model is a better choice, if possible.
One should also keep in mind that despite the cavity radiation being resonant with (ro)vibrational transitions, the cavity-induced changes in the electronic structure might be significant \cite{Bonini2022,23Sz,Horak2024}.

Finally, a unique and important feature of the cavity setting is that because a single mode of the cavity radiation field can simultaneously interact with different systems within the cavity, the radiation field can act as an indirect interaction between different molecular or atomic species \cite{21SzBaHaVi,Fleischer_2019,cavity_Zhong_AngChem_2016,cavity_Muallem_JPCL_2016,Xiang_2020,Takele2020}, and lead to so-called collective effects \cite{Ebbesen_review_2016,Herrera_2020,Scholes2021,22Ce}.
Collective effects play a central role in polaritonic chemistry, for example, they are responsible for the well-known $\sqrt{n}$ scaling of the light-matter coupling strength when $n$ molecules interact with the cavity mode.
The collective states that are formed are coherent superpositions of multiple molecular and photonic states.
Although it is debated whether quantum coherence on a mesoscopic scale can indeed be realized \cite{Sidler2022}, the existence of collective states has been a key factor in considering and describing the physicochemical properties and reactions of vibropolaritonic systems \cite{Herrera2017,Vendrell2018_b,Xiang2018,18HeSp,CamposGonzalezAngulo_2020,Du2022,Gmez2023,Lindoy2024}.

In this work we employ realistic rovibrational molecular models, originally developed for high-resolution molecular spectroscopy, to gain a rotationally resolved insight into the collective effects and intermolecular processes arising when molecules in the gas phase are confined in a resonant IR cavity.
On the other hand, including into the simulations the very large number of molecules used in currently existing gas-phase experiments is practically impossible with the theory outlined below.
Therefore, the idea here is that cavity-molecule interactions are simulated using a single- or few-molecule model with an increased effective coupling, representing the experimental number of molecules, aiming to capture all non-collective effects, as well as to some extent the most important collective effects (see also Ref. \cite{Horak2024} for the justification of this approach).

The theoretical approach is detailed in the next section, which is followed by applications on HCl and H$_2$O molecules.
The energetics, spectral and dynamic properties are investigated, with emphasys on the advantages and importance of using rotationally resolved molecular models.
The simulations on HCl and H$_2$O use parameters which do not resemble any existing experimental setup, but which are chosen to conveniently showcase the capabilities of the theory introduced below.
The paper is closed with simulation results on CH$_4$, using parameters representing the setup of the experimental works of Refs. \cite{Wright2023,Wright2023_b}.

\section{\label{Theory}Theory}
The starting point of our approach is the versatile theoretical framework developed in Ref. \cite{23Sz}, in which the exact quantum treatment of molecular rotation, as well as the cavity-induced changes in electronic structure are incorporated, meanwhile molecular vibrations can be treated on a desired level of sophistication.
The protocol of Ref. \cite{23Sz} is composed of two steps: first, the field-free molecular eigenstates are obtained within a chosen molecular model, and second, the molecule + radiation mode(s) Hamiltonian is constructed using the $\vert N\rangle \vert \Psi^{nJM}\rangle$ direct-product basis (for simplicity assuming a single radiation mode), where the $\vert\Psi^{nJM}\rangle$ field-free rovibrational eigenstates satisfy $ \hat{H}_{{\rm rovib}}\vert\Psi^{nJM}\rangle=E^{nJ}\vert\Psi^{nJM}\rangle$, where $J$ and $M$ are the rotational angular momentum and its projection onto the space-fixed z-axis, respectively, $n$ is all other quantum numbers uniquely defining the rovibrational states, and $\vert N\rangle$ is a photon number state of the cavity radiation.

For multiple, $N_{\rm mol}$ molecules interacting with a radiation mode,
but not directly with eachother, i.e., assuming an ideal gas scenario,
following the approach of Ref. \cite{23Sz} leads to the Hamiltonian
\begin{equation}
\begin{aligned}
\hat{H}&=
\sum_i^{N_{\rm mol}} \hat{H}_{{\rm rovib}}^{(i)}
+\hbar\omega_{{\rm c}}\hat{a}_{{\rm c}}^{\dagger}\hat{a}_{{\rm c}}
 -\frac{g}{ea_{0}} \sum_i^{N_{\rm mol}} \mathbf{e} \hat{\mathbf{\upmu}}_0^{(i)}(\hat{a}_{{\rm c}}^{\dagger}+\hat{a}_{{\rm c}})\\
&-\frac{(g/ea_{0})^2}{2} \sum_i^{N_{\rm mol}} \mathbf{e}\hat{\mathbf{\alpha}}^{(i)}\mathbf{e}(\hat{a}_{{\rm c}}^{\dagger}+\hat{a}_{{\rm c}})(\hat{a}_{{\rm c}}^{\dagger}+\hat{a}_{{\rm c}})
 + \frac{(g/ea_{0})^2}{\hbar\omega_{{\rm c}}}(\sum_i^{N_{\rm mol}} \mathbf{e}\hat{\mathbf{\upmu}}_0^{(i)})^2\\
&+O(g^3),
\end{aligned}
\label{eq:hamiltonian_full}
\end{equation}
where the quantity $g=e a_{0} \sqrt{\hbar\omega_{{\rm c}}/(2 \varepsilon_{0}V)}$ represents the coupling strength between a single photon electric field and the atomic unit of the dipole moment, $\hat{H}_{{\rm rovib}}^{(i)}$ is the field-free rovibrational Hamiltonian of the \textit{i}th molecule, while $\hat{\mathbf{\upmu}}_0^{(i)}$ and $\hat{\mathbf{\alpha}}^{(i)}$ are the permanent dipole and polarizability of the $i$th molecule, respectively, both being a function of the nuclear coordinates.
Note that in the last term of Eq. (\ref{eq:hamiltonian_full}), which is called the self-dipole or dipole self-energy (DSE) term, the approximation $\langle \mathbf{\upmu}^2 \rangle \approx \langle \mathbf{\upmu} \rangle^2$ is made, where $\langle ... \rangle$ denotes the expectation value in the electronic ground state.
In words, the expectation value of the square dipole is approximated with the square of the dipole expectation value, an approach also used in previous works \cite{23Sz,Yu2024}.
This approximation can be convenient and necessary from a practical perspective, because nuclear-coordinate dependent DSE surfaces are not available in the literature, however, caution is needed, as pointed out below.

The direct product basis functions used to construct the matrix representation of the Hamiltonian in Eq. (\ref{eq:hamiltonian_full}) are
 \begin{equation}
 \Bra{N} \Bra{\Psi^{n^{(1)}J^{(1)}M^{(1)}}} \otimes \Bra{\Psi^{n^{(2)}J^{(2)}M^{(2)}}}  \otimes \cdots \otimes \Bra{\Psi^{n^{(N_{\rm mol})}J^{(N_{\rm mol})}M^{(N_{\rm mol})}}}.
 \label{eq:dp_basis_general}
 \end{equation}

For simplicity the following formulae are given for the case of two molecules interacting with the cavity mode.
Extending to a larger number of molecules and cavity modes is in principal straightforward.
In the case of two molecules one obtains for the first term of the Hamiltonian
\begin{footnotesize}
 \begin{equation}
 \begin{aligned}
 &\Bra{N}\red{\Bra{\Psi^{nJM}}}\blue{\Bra{\Psi^{nJM}}} ( \hat{H}_{{\rm rovib}}^{(1)} + \hat{H}_{{\rm rovib}}^{(2)} ) \Ket{N'}\red{\Ket{\Psi^{n'J'M'}}}\blue{\Ket{\Psi^{n'J'M'}}} = \\
 & \delta_{NN'} ( \red{E^{nJ}}\red{\delta_{n n'}\delta_{J J'}\delta_{M M'}} \blue{\delta_{n n'}\delta_{J J'}\delta_{M M'}} +\blue{E^{nJ}}\red{\delta_{n n'}\delta_{J J'}\delta_{M M'}} \blue{\delta_{n n'}\delta_{J J'}\delta_{M M'}} ),
 \end{aligned}
 \label{eq:fieldfree_matrix_general}
 \end{equation}
\end{footnotesize}
\noindent
where for clarity we used red[blue] coloring instead of the (1)[(2)] superscripts to distinguish the molecules in the basis functions.
The matrix elements of the second term in the Hamiltonian are
\begin{footnotesize}
 \begin{equation}
 \Bra{N}\red{\Bra{\Psi^{nJM}}}\blue{\Bra{\Psi^{nJM}}} \hbar\omega_{{\rm c}}\hat{a}_{{\rm c}}^{\dagger}\hat{a}_{{\rm c}} \Ket{N'}\red{\Ket{\Psi^{n'J'M'}}}\blue{\Ket{\Psi^{n'J'M'}}} = \hbar\omega_{{\rm c}} N \delta_{NN'} \red{\delta_{n n'}\delta_{J J'}\delta_{M M'}} \blue{\delta_{n n'}\delta_{J J'}\delta_{M M'}}.
 \label{eq:radiationmode_matrix_general}
 \end{equation}
\end{footnotesize}
\noindent
For the dipole interaction terms one gets
\begin{footnotesize}
 \begin{equation}
 \begin{aligned}
 - \frac{g}{ea_{0}}  &\Bra{N}\red{\Bra{\Psi^{nJM}}}\blue{\Bra{\Psi^{nJM}}} (\hat{\mu}_z^{\rm (1),SF}+\hat{\mu}_z^{\rm (2),SF})(\hat{a}_{{\rm c}}^{\dagger}+\hat{a}_{{\rm c}}) \Ket{N'}\red{\Ket{\Psi^{n'J'M'}}}\blue{\Ket{\Psi^{n'J'M'}}} = \\
 - \frac{g}{ea_{0}} &(
  \Bra{N}\hat{a}_{{\rm c}}^{\dagger} + \hat{a}_{\rm c} \Ket{N'} \red{\Bra{\Psi^{nJM}}} \hat{\mu}_z^{\rm (1),SF}\red{\Ket{\Psi^{n'J'M'}}} \blue{\Braket{\Psi^{nJM}|\Psi^{n'J'M'}}} + \\
  &\Bra{N}\hat{a}_{{\rm c}}^{\dagger} + \hat{a}_{\rm c} \Ket{N'} \red{\Braket{\Psi^{nJM}|\Psi^{n'J'M'}}} \blue{\Bra{\Psi^{nJM}}} \hat{\mu}_z^{\rm (2),SF}\blue{\Ket{\Psi^{n'J'M''}}}) =  \\
 - \frac{g}{ea_{0}} &( (\sqrt{N'+1}\delta_{NN'+1} + \sqrt{N'}\delta_{NN'-1}) \times \\
 &  \red{\Bra{\Psi^{nJM}}} \hat{\mu}_z^{\rm (1),SF}\red{\Ket{\Psi^{n'J'M'}}} \blue{\delta_{n n'}\delta_{J J'}\delta_{M M'}} + \blue{\Bra{\Psi^{nJM}}} \hat{\mu}_z^{\rm (2),SF}\blue{\Ket{\Psi^{n'J'M'}}} \red{\delta_{n n'}\delta_{J J'}\delta_{M M'}} ),
 \end{aligned}
 \label{eq:dipole_matrix_general}
 \end{equation}
\end{footnotesize}
\noindent
where the superscripts SF represent that the dipole components are those along the polarization vector of the cavity mode, i.e., along the space-fixed (SF) z-axis.
In a similar fashion, the matrix elements of the polarizability interaction take the form
\begin{footnotesize}
 \begin{equation}
 \begin{aligned}
 - \frac{(g/ea_{0})^2}{2}  &\Bra{N}\red{\Bra{\Psi^{nJM}}}\blue{\Bra{\Psi^{nJM}}} (\hat{\alpha}_{zz}^{\rm (1),SF}+\hat{\alpha}_{zz}^{\rm (2),SF})(\hat{a}_{{\rm c}}^{\dagger}+\hat{a}_{{\rm c}})(\hat{a}_{{\rm c}}^{\dagger}+\hat{a}_{{\rm c}}) \Ket{N'}\red{\Ket{\Psi^{n'J'M'}}}\blue{\Ket{\Psi^{n'J'M'}}} = \\
       &-\frac{(g/ea_{0})^2}{2} \Big( \sqrt{(N'+1)(N'+2)}\delta_{N,N'+2}+(2N'+1)\delta_{N,N'}+\sqrt{N'(N'-1)}\delta_{N,N'-2}\Big)\times \\
        &  ( \red{\Bra{\Psi^{nJM}}} \hat{\alpha}_{zz}^{\rm (1),SF} \red{\Ket{\Psi^{n'J'M'}}} \blue{\delta_{n n'}\delta_{J J'}\delta_{M M'}} + \blue{\Bra{\Psi^{nJM}}} \hat{\alpha}_{zz}^{\rm (2),SF}\blue{\Ket{\Psi^{n'J'M'}}} \red{\delta_{n n'}\delta_{J J'}\delta_{M M'}} ).
 \end{aligned}
 \label{eq:polarizability_matrix_general}
 \end{equation}
\end{footnotesize}
\noindent
As for the dipole self-energy term, one can write
\begin{footnotesize}
 \begin{equation}
 \begin{aligned}
 \frac{(g/ea_{0})^2}{\hbar \omega_c}  &\Bra{N}\red{\Bra{\Psi^{nJM}}}\blue{\Bra{\Psi^{nJM}}} ({\hat{\mu}_{z}^{\rm (1),SF}}\hat{\mu}_{z}^{\rm (1),SF}+{\hat{\mu}_{z}^{\rm (2),SF}}\hat{\mu}_{z}^{\rm (2),SF}+2\hat{\mu}_{z}^{\rm (1),SF}\hat{\mu}_{z}^{\rm (2),SF}) \Ket{N'}\red{\Ket{\Psi^{n'J'M'}}}\blue{\Ket{\Psi^{n'J'M'}}} = \\
 \frac{(g/ea_{0})^2}{\hbar \omega_c} \delta_{N,N'} &
 ( \red{\Bra{\Psi^{nJM}}} {\hat{\mu}_{z}^{\rm (1),SF}}\hat{\mu}_{z}^{\rm (1),SF} \red{\Ket{\Psi^{n'J'M'}}} \blue{\delta_{n n'}\delta_{J J'}\delta_{M M'}} +
 \blue{\Bra{\Psi^{nJM}}} {\hat{\mu}_{z}^{\rm (2),SF}}\hat{\mu}_{z}^{\rm (2),SF} \blue{\Ket{\Psi^{n'J'M'}}} \red{\delta_{n n'}\delta_{J J'}\delta_{M M'}} + \\
 & 2 \red{\Bra{\Psi^{nJM}}} {\hat{\mu}_{z}^{\rm (1),SF}} \red{\Ket{\Psi^{n'J'M'}}} \blue{\Bra{\Psi^{nJM}}} \hat{\mu}_{z}^{\rm (2),SF} \blue{\Ket{\Psi^{n'J'M'}}} ),
 \end{aligned}
 \label{eq:selfdipole_matrix_general}
 \end{equation}
\end{footnotesize}

For the simplest case of diatomic molecules, the SF components of the dipole are transformed to the body-fixed (BF) components by $\hat{\mu}_z^{\rm (i),SF}=\hat{\mu}_z^{\rm (i),BF}{\rm cos}(\theta^{\rm (i)})$, where $\theta$ is the rotational coordinate between the SF and BF z-axes, and the polarizability interaction terms $\mathbf{e}\hat{\mathbf{\alpha}}^{(i)}\mathbf{e}$ for a linearly polarized cavity mode can be expressed as $\mathbf{e}\hat{\mathbf{\alpha}}^{(i)}\mathbf{e}=\alpha_\parallel {\rm cos}^2(\theta^{(i)}) + \alpha_\perp {\rm sin}^2(\theta^{(i)})=(\alpha_\parallel - \alpha_\perp) {\rm cos}^2(\theta^{(i)}) + \alpha_\perp$ with the $\alpha_\parallel$ parallel and $\alpha_\perp$ perpendicular components of the polarizability.
Note that in this work the DSE term is approximated as the square of the dipole expectation values, which gives $\mu^{\rm SF}_z \mu^{\rm SF}_z = \mu^{\rm BF}_z \mu^{\rm BF}_z {\rm cos}^2(\theta)$, but without this approximation one would get $\langle \mu^{\rm SF}_z \mu^{\rm SF}_z \rangle = \langle \mu^{\rm BF}_z \mu^{\rm BF}_z \rangle {\rm cos}^2(\theta) + \langle \mu^{\rm BF}_\perp \mu^{\rm BF}_\perp \rangle {\rm sin}^2(\theta) = ( \langle \mu^{\rm BF}_z \mu^{\rm BF}_z \rangle - \langle \mu^{\rm BF}_\perp  \mu^{\rm BF}_\perp \rangle ) {\rm cos}^2(\theta) + \langle \mu^{\rm BF}_\perp \mu^{\rm BF}_\perp \rangle$, where $\perp$ indicates the dipole component parallel to the molecular axis, and, as before, $\langle ... \rangle$ stands for the expectation value in the ground electronic state.
Therefore, the exact treatment does not change the selection rules for the interaction, but adds a constant shift to the energy levels and uses $\langle \mu^{\rm BF}_z \mu^{\rm BF}_z - \mu^{\rm BF}_\perp  \mu^{\rm BF}_\perp \rangle$ instead of $\langle \mu^{\rm BF}_z \rangle \langle \mu^{\rm BF}_z \rangle$.
Unfortunately, computing the expectation value of the dipole square is not trivial \cite{Borges2024,24Fabri_SDE}, but preliminary computations by one of the Authors of Ref. \cite{24Fabri_SDE} on HCl near the equilibrium distance indicate that the approximation for the DSE in this work underestimates the effect of the DSE term by at least a factor of two.
This should be kept in mind regarding the numerical results presented below, nonetheless, this does not affect the main messages to be conveyed in this work.
In light of the above, and using $\ket{nJM}=\ket{n}\ket{JM}$ field-free basis functions expressed as a product of a vibrational wave function $\ket{n}$ and a rotational wave function $\ket{JM}$ (which are the spherical harmonics if coordinate representation is used \cite{06BuJe,88Zare}), Eq. (\ref{eq:dipole_matrix_general}) reads
\begin{footnotesize}
 \begin{equation}
 \begin{aligned}
 - \frac{g}{ea_{0}}  &\Bra{N}\red{\Bra{nJM}}\blue{\Bra{nJM}} (\hat{\mu}_z^{\rm (1),SF}+\hat{\mu}_z^{\rm (2),SF})(\hat{a}_{{\rm c}}^{\dagger}+\hat{a}_{{\rm c}}) \Ket{N'}\red{\Ket{n'J'M'}}\blue{\Ket{n'J'M'}} = \\
 - \frac{g}{ea_{0}} &( (\sqrt{N'+1}\delta_{NN'+1} + \sqrt{N'}\delta_{NN'-1}) \times \\
 & ( \red{\Bra{n}} \hat{\mu}_z^{\rm (1),BF}\red{\Ket{n'}}\red{\Bra{JM}} {\rm cos}(\theta^{\rm (1)})\red{\Ket{J'M'}} \blue{\delta_{J J'}\delta_{M M'}} + \\
 &  \blue{\Bra{n}} \hat{\mu}_z^{\rm (2),BF}\blue{\Ket{n'}} \blue{\Bra{JM}} {\rm cos}(\theta^{\rm (2)})\blue{\Ket{J'M'}} \red{\delta_{J J'}\delta_{M M'}} ).
 \end{aligned}
 \label{eq:diatomic_dipole_1}
 \end{equation}
\end{footnotesize}
Eq. (\ref{eq:polarizability_matrix_general}) for diatomics is
\begin{footnotesize}
 \begin{equation}
 \begin{aligned}
 - \frac{(g/ea_{0})^2}{2}  &\Bra{N}\red{\Bra{nJM}}\blue{\Bra{nJM}} (\hat{\alpha}_{zz}^{\rm (1),SF}+\hat{\alpha}_{zz}^{\rm (2),SF})(\hat{a}_{{\rm c}}^{\dagger}+\hat{a}_{{\rm c}})(\hat{a}_{{\rm c}}^{\dagger}+\hat{a}_{{\rm c}}) \Ket{N'}\red{\Ket{n'J'M'}}\blue{\Ket{n'J'M'}} = \\
       &-\frac{(g/ea_{0})^2}{2} \Big( \sqrt{(N'+1)(N'+2)}\delta_{N,N'+2}+(2N'+1)\delta_{N,N'}+\sqrt{N'(N'-1)}\delta_{N,N'-2}\Big)\times \\
  \Big( & \big[ \red{\Bra{n}} \hat{\alpha}_\parallel^{\rm (1)} - \hat{\alpha}_\perp^{\rm (1)} \red{\Ket{n'}}\red{\Bra{JM}} {\rm cos}^2(\theta^{\rm (1)})\red{\Ket{J'M'}} + \red{\Bra{n}} \hat{\alpha}_\perp^{\rm (1)}\red{\Ket{n}} \red{\delta_{J J'}\delta_{M M'}} \big] \blue{\delta_{n n'}\delta_{J J'}\delta_{M M'}} + \\
 & \big[ \blue{\Bra{n}} \hat{\alpha}_\parallel^{\rm (2)} - \hat{\alpha}_\perp^{\rm (2)} \blue{\Ket{n'}}\blue{\Bra{JM}} {\rm cos}^2(\theta^{\rm (2)})\blue{\Ket{J'M'}} + \blue{\Bra{n}} \hat{\alpha}_\perp^{\rm (2)}\blue{\Ket{n}} \blue{\delta_{J J'}\delta_{M M'}} \big] \red{\delta_{n n'}\delta_{J J'}\delta_{M M'}} \Big),
 \end{aligned}
 \label{eq:diatomic_polarizability_1}
 \end{equation}
\end{footnotesize}
and finally Eq. (\ref{eq:selfdipole_matrix_general}) reads
\begin{footnotesize}
 \begin{equation}
 \begin{aligned}
 \frac{(g/ea_{0})^2}{\hbar \omega_c}  &\Bra{N}\red{\Bra{nJM}}\blue{\Bra{nJM}} ({\hat{\mu}_{z}^{\rm (1),SF}}\hat{\mu}_{z}^{\rm (1),SF}+{\hat{\mu}_{z}^{\rm (2),SF}}\hat{\mu}_{z}^{\rm (2),SF}+2\hat{\mu}_{z}^{\rm (1),SF}\hat{\mu}_{z}^{\rm (2),SF}) \Ket{N'}\red{\Ket{n'J'M'}}\blue{\Ket{n'J'M'}} = \\
 \frac{(g/ea_{0})^2}{\hbar \omega_c} \delta_{N,N'} &
 ( \red{\Bra{n}} {\hat{\mu}_{z}^{\rm (1),BF}}\hat{\mu}_{z}^{\rm (1),BF} \red{\Ket{n'}} \red{\Bra{JM}} {\rm cos}^2(\theta^{\rm (1)}) \red{\Ket{J'M'}} \blue{\delta_{n n'}\delta_{J J'}\delta_{M M'}} + \\
 & \blue{\Bra{n}} {\hat{\mu}_{z}^{\rm (2),BF}}\hat{\mu}_{z}^{\rm (2),BF} \blue{\Ket{n'}} \blue{\Bra{JM}} {\rm cos}^2(\theta^{\rm (1)}) \blue{\Ket{J'M'}} \red{\delta_{n n'}\delta_{J J'}\delta_{M M'}} + \\
 & 2 \red{\Bra{n}} {\hat{\mu}_{z}^{\rm (1),BF}} \red{\Ket{n'}} \blue{\Bra{n}} \hat{\mu}_{z}^{\rm (2),BF} \blue{\Ket{n'}} \red{\Bra{JM}} {\rm cos}(\theta^{\rm (1)}) \red{\Ket{J'M'}} \blue{\Bra{JM}} {\rm cos}(\theta^{\rm (2)}) \blue{\Ket{J'M'}} ),
 \end{aligned}
 \label{eq:diatomic_selfdipole_1}
 \end{equation}
\end{footnotesize}
In \cref{eq:diatomic_dipole_1,eq:diatomic_polarizability_1,eq:diatomic_selfdipole_1} the rotational matrix elements can be evaluated using three-j symbols \cite{88Zare,06BuJe} as
\begin{footnotesize}
 \begin{equation}
 \begin{aligned}
 & \Bra{JM} {\rm cos}(\theta)\Ket{J'M'} = \sqrt{(2J+1)(2J'+1)} (-1)^{M'} \threej{J}{1}{J'}{M}{0}{-M'}\threej{J}{1}{J'}{0}{0}{0} \\
 & \Bra{JM} {\rm cos}^2(\theta)\Ket{J'M'} = \frac{1}{3}\delta_{JJ'}\delta_{MM'}+ \frac{2}{3}\sqrt{(2J+1)(2J'+1)}(-1)^{M'} \threej{J}{2}{J'}{M}{0}{-M'} \threej{J}{2}{J'}{0}{0}{0}. \\
 \end{aligned}
 \label{eq:threej_diatomic}
 \end{equation}
\end{footnotesize}

For two polyatomic molecules the formulae are more involved, but the general expressions can be obtained with reasonable effort based on the equations above and those given in Ref. \cite{23Sz}.

Prior to any numerical computations, inspecting the non-zero matrix elements of the different terms in the Hamiltonian above can reveal the possible channels among which the various molecules and degrees of freedom can interact, i.e., based on the equations above, one can identify the selection rules of the interactions governing cavity-mediated energy transfer and collective polaritonic state formation.
For example, the dipole interaction in \cref{eq:diatomic_dipole_1} requires that only one molecule changes its state with $\Delta J = \pm 1$, $\Delta M = 0$, and $\Delta N = \pm 1$ for the cavity, given that the $\Bra{n}\hat{\mu}_z^{\rm BF}\Ket{n'}$ transition dipole for that molecule is not zero.
Cavity-mediated dipole interaction between the two molecules is thus achieved as a two-step process.
For example, in a time-dependent picture for a resonant interaction to occur, initially one molecule emits(absorbs) a photon while changing state, then as the second step the other molecule absorbs(emits) a photon and changes state.
The polarizability interaction between two molecules in \cref{eq:diatomic_polarizability_1} is also a two-step process with the selection rules for each step being $\Delta J = 0, \pm 2$, $\Delta M = 0$, $\Bra{n} \hat{\alpha}_\parallel - \hat{\alpha}_\perp \Ket{n'} \ne 0$ for one molecule and no change for the other and $\Delta N = 0, \pm 2$ for the cavity.
The self-dipole interaction between the molecules can be realized either as a one- or two-step process.
The two-step process has for each step the selection rules $\Delta J = 0, \pm 2$, $\Delta M = 0$ and $\Bra{n}\hat{\mu}_z^{\rm BF}\hat{\mu}_z^{\rm BF}\Ket{n'}\ne 0$ for one molecule and no change for the other, with $\Delta N =0$ for the cavity.
The one-step process requires for both molecules $\Delta J = \pm 1$, $\Delta M = 0$ and $\Bra{n}\hat{\mu}_z^{\rm BF}\Ket{n'} \ne 0$, with $\Delta N =0$ for the cavity.
For polyatomics the situaition is very similar, but the general selection rules are $\Delta J = 0, \pm 1$, $\Delta M = 0$ for the dipole interaction (and one-step self-dipole interaction), $\Delta J = 0, \pm 1, \pm 2$, $\Delta M = 0$ for the polarizability interaction (and two-step self-dipole interaction), and there are a larger number of different transition matrix elements, some of which, if zero, can lead to more strict selection rules.
The selection rules above can be useful to interpret the light-dressed spectra of the polaritonic systems and to understand the mechanisms governing their quantum dynamics.

Once the matrix representation of the Hamiltonian in Eq. (\ref{eq:hamiltonian_full}) is obtained using the basis functions of Eq. (\ref{eq:dp_basis_general}), the Hamiltonian matrix is diagonalized to obtain the $E^{\rm pol} _k $ polaritonic energies and $\ket{\psi ^{\rm pol} _k}$ wavefunctions.
The molecular absorbtion spectra are computed using the simplest formulation of light-dressed spectroscopy \cite{19SzCsHaVi,PUILSXV} as implemented in the context of molecular polaritonics \cite{18SzHaCsCe_a,20SzHaVi,21SzBaHaVi}, i.e., the $T_{l \leftarrow k}$ transition amplitude between the $k$th and $l$th polaritonic state is evaluated as $T_{l \leftarrow k} \propto \langle \psi^{\rm pol}_l \vert \sum_i^{N_{\rm mol}} \hat{\mathbf{\upmu}}_0^{(i)}\hat{\bm{e}}^{(p)} \vert \psi^{\rm pol}_k \rangle \delta(E^{\rm pol}_l - E^{\rm pol}_k \pm \hbar \omega_p)$, where $\hat{\bm{e}}^{(p)}$ and $\hbar \omega_p$ are the polarization vector and photon energy of the probe pulse, respectively.
In a similar fashion, cavity absorbtion is computed by $T^{\rm cav}_{l \leftarrow k} \propto \langle \psi^{\rm pol}_l \vert (\hat{a}_{{\rm c}}^{\dagger}+\hat{a}_{{\rm c}}) \vert \psi^{\rm pol}_k \rangle \delta(E^{\rm pol}_l - E^{\rm pol}_k \pm \hbar \omega_p)$.

The quantum dynamics simulations were carried out by expanding the $\ket{\Psi(t)}$ time-dependent wave function in the basis of the polaritonic eigenstates, $\ket{\Psi(t)}=\sum_k c_k(t) \ket{\psi^{\rm pol}_k}$, and solving the time-dependent Schrödinger equation.
For non-Hermitian dynamics, the Hamiltonian was expanded with the term $-i \frac{\gamma_c}{2} \hat{a}_{{\rm c}}^{\dagger}\hat{a}_{{\rm c}}$, where $\gamma_c$ is the cavity decay rate \cite{Visser1995}, which is a computationally efficient alternative to using the density matrix and the Lindblad equations \cite{Manzano2020}, given that no
decay from higher-lying states and no pumping is present
, and the ground state populations are not of interest \cite{24FaCsaHaCe}.
Some simulations were also repeated using the Lindblad formalism to test the validity of the non-Hermitian approach.

In order to obtain the $\vert \Psi^{nJM}\rangle$ field-free rovibrational eigenstates, numerically exact variational models based on the discrete variable representation (DVR) \cite{00LiCa} were used.
Solving the rovibrational problem for the diatomic HCl molecules in DVR is straightforward and was done with an in-house code, adopting the potential energy curve of Ref. \cite{Domcke1985} and the atomic masses $m_{\rm H} = 1.00784$ u and $m_{\rm  Cl}=34.96885$ u.
The dipole and polarizability matrix elements were computed using the curves from Refs. \cite{Harrison2008} and \cite{Maroulis1998}, respectively.
The field-free computations involving H$_2$O were carried out in an identical manner to those detailed in Ref. \cite{23Sz} for the variational model.
In short, the 6-D rovibrational problem was solved with the D2FOPI code \cite{10SzCsCz}, using the potential energy surface of Ref. \cite{18PoKyZoTe}.
The dipole and polarizability matrix elements were generated using the surfaces from Refs. \cite{Lodi2011} and \cite{Avila2005}, respectively.
All computed field-free rovibrational energies are converged to within 0.01 cm$^{-1}$.
For CH$_4$, the rovibrational energy levels were adopted from \cite{Kefala2024}, vibrational transition dipoles were taken from \cite{Yurchenko2013}, and rovibrational transition dipoles were computed using the formulae of Ref. \cite{23Sz}, assuming an exact separation between vibration and rotation.
Given the accuracy of the PESs near the equilibrium structure, the computed rovibrational transition energies differ from their experimental counterparts by a few cm${^{-1}}$ for HCl \cite{Domcke1985}, a few 0.01 cm$^{-1}$ for H$_2$O \cite{18PoKyZoTe}.
The error of the energy levels adopted for CH$_4$ are less than a few 0.001 cm$^{-1}$ \cite{Kefala2024}.
Because $\Delta M = 0$ holds for all interactions, we omit this quantum number in the results shown below, and use the notations $\ket{v J}$ and $\ket{(v_1 v_2 v_3)(J K_a K_c)}$ for the rovibrational eigenstates of HCl and H$_2$O, respectively, where the quantum numbers have their usual meaning \cite{06BuJe,10MaFaSzCz,10CsMaSzLo}.
For both HCl and H$_2$O molecules ten field-free eigenstates were included in the basis of Eq. (\ref{eq:dp_basis_general}), and a maximal photon number of three was used.
For HCl this included the vibrational ground state and first fundamental with rotational states up to $J=4$.
For H$_2$O the vibrational ground state, the bending fundamental, and all rotational states up to $J=2$ were included in the basis.
For CH$_4$ either two or four eigenstates were considered, depending on the experimental setup to be simulated, and up to three cavity modes were included with up to one photon in each mode.

\section{Results and discussion}

%\subsection{\label{polaritons}Polaritons}

\subsection{\label{sec:energetics}Energetics of HCl and H$_2$O molecules in an IR cavity}
\begin{figure}[!ht]
\centering \includegraphics[width=0.95\textwidth]{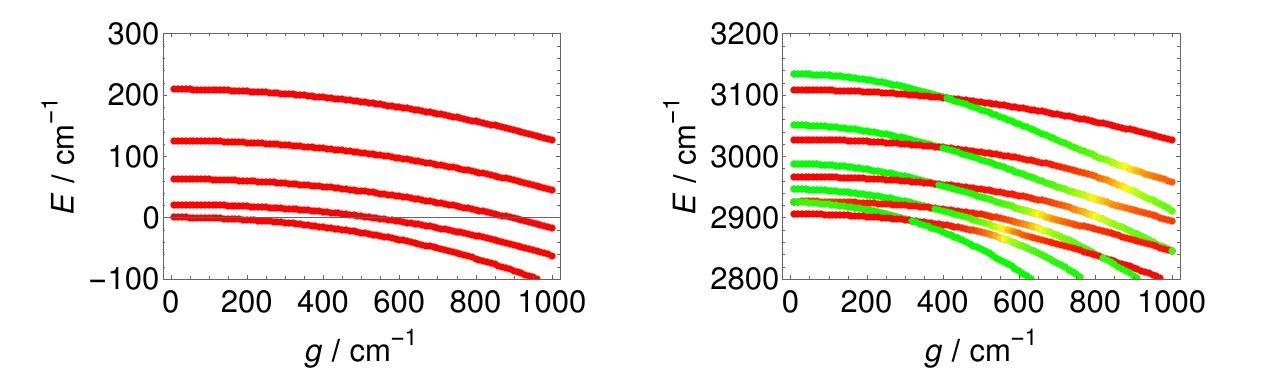}
\centering \includegraphics[width=0.95\textwidth]{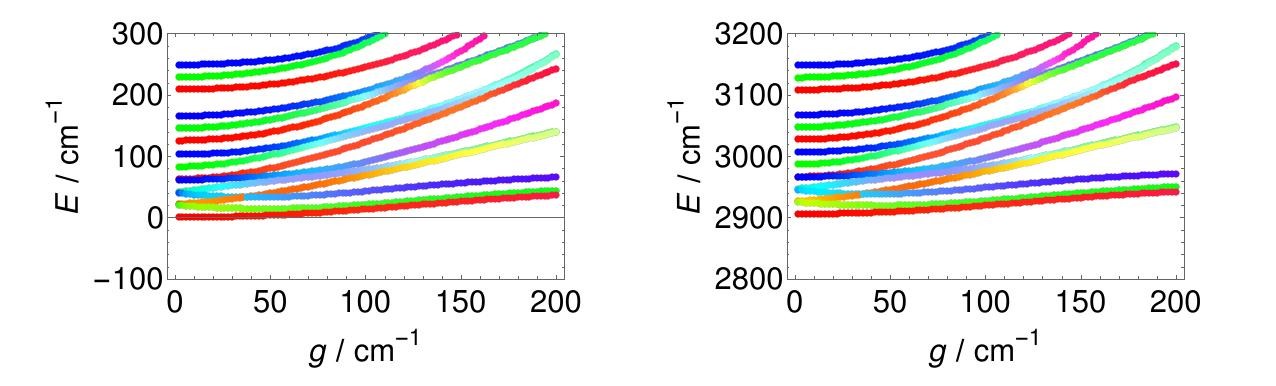}
\caption{Energy levels of the ``HCl + IR cavity mode" system as a function of the $g$ coupling strengh. Upper row: $\hbar \omega_c = 2925$ cm$^{-1}$, lower row: $\hbar \omega_c = 20$ cm$^{-1}$.
The colors of the lines represent their character: red indicates zero expectation value for the photon number, while green and blue represent one and two photon expectation values, respectively. Yellow, purple, teal, etc. indicate a mixture of various photonic and material excitations.
}

\label{fig:HCl_gfugges}
\end{figure}

First, we briefly examine the characteristics of the system when a single HCl molecule interacts with a cavity mode resonant to its vibrational fundamental.
As can be seen in Fig. \ref{fig:HCl_gfugges} and also observed in Ref. \cite{23Sz}, the formation of polaritonic states is not guaranteed by setting the cavity photon energy to be resonant with an optically allowed transition, such as $\ket{1 1} \leftarrow \ket{0 0}$ at 2925.8 cm$^{-1}$, because the energy levels can shift with respect to their field-free values in the cavity due to polarization.
In the simulations with $\hbar\omega_c=2925$ cm$^{-1}$, the energy levels all shift down with increasing $g$, and the states having a photon in the cavity (green curves in the upper right panel of Fig \ref{fig:HCl_gfugges}) shift more rapidly, due to the $(2N'+1)\delta_{N N'}$ term appearing in the one-molecule analog of Eq. \ref{eq:diatomic_polarizability_1}.
The vibrational transition dipoles, i.e., the matrixelements $\bra{n} \mu_z^{\rm (BF)} \ket{n'}, n\ne n'$ are fairly small for HCl ($\bra{0} \mu_z^{\rm (BF)} \ket{1}\approx0.027$ au), therefore, when $\hbar\omega_c=2925$ cm$^{-1}$, $g$ values larger than 500 cm$^{-1}$ are needed for polariton formation; the exact value depending on the $J$ and $J'=J\pm1$ rotational quantum numbers, because the energy levels of the polariton forming states with different $J$ shift close to each other at different $g$ values.
In the low energy rotation-only region (upper left panel of Fig. \ref{fig:HCl_gfugges}) no polariton formation can be seen, as expected.
The $\bra{n} \mu_z^{\rm (BF)} \ket{n}$ permanent dipoles of HCl, however, are considerable ($\bra{0} \mu_z^{\rm (BF)} \ket{0}\approx \bra{1} \mu_z^{\rm (BF)} \ket{1} \approx 0.43$ au), leading to efficient rotational polariton formation even at small $g$ values. For $\hbar\omega_c=20$ cm$^{-1}$ the lower row of Fig. \ref{fig:HCl_gfugges} demonstrates that rotational polaritons are formed both in the low energy rotation-only region and near the vibrational fundamental.
The efficient rotational polariton formation might make HCl a promising candidate for gas-phase polaritonic experiments in cm scale Fabry-Pérot cavities \cite{Wright2023_b}.

\begin{figure}[!ht]
\centering \includegraphics[width=0.35\textwidth]{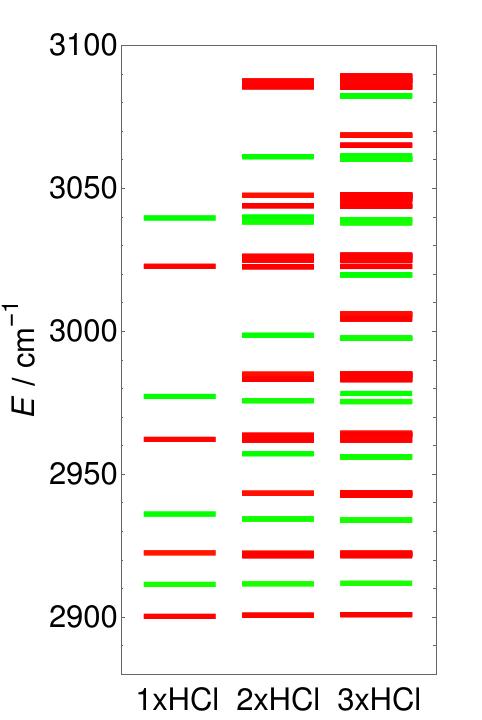}
\centering \includegraphics[width=0.35\textwidth]{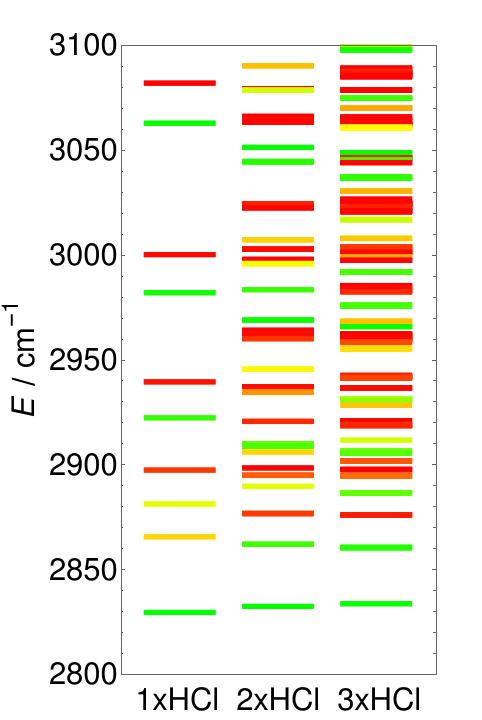}
\caption{Energy levels of the ``$n\times$HCl + IR cavity mode" system for $\hbar \omega_c = 2925$ cm$^{-1}$. Left panel: $g = 200/\sqrt{N_{\rm mol}}$ cm$^{-1}$, right panel: $g = 550/\sqrt{N_{\rm mol}}$ cm$^{-1}$. The colors of the lines represent their character: red indicates zero expectation value for the photon number, while green represents one photon expectation value. Yellow indicates a mixture of photonic and material excitations.
}
\label{fig:3xHCl_2925}
\end{figure}

The effects of increasing the number of HCl molecules in the cavity
%will be examined in detail in the context of spectroscopy and time-dependent dynamics below, here we only note that increasing the number of molecules
allows for additional combinations in the excitation of the different degress of freedom in the different molecules, which ultimately leads to a very dense and complex energy spectrum of the whole system, as seen in Fig. \ref{fig:3xHCl_2925}.
Fig. \ref{fig:3xHCl_2925} also demonstrates that collective effects are in place. Despite the $1/\sqrt{N_{\rm mol}}$ scaling of the $g$ coupling strength, not only does the energy spectrum get more dense with increasing molecule number, but also the position of many energy levels are shifted.
These shifts are moderate at the lower coupling strength of $g=200$ cm$^{-1}$ shown in the left panel of Fig. \ref{fig:3xHCl_2925}, but are very pronounced at the $g=550$ cm$^{-1}$ coupling strength, at which polaritons are also formed, see right panel of Fig. \ref{fig:3xHCl_2925}.

%In contrast, collective effects are not so apparent in the energetics for H$_2$O molecules, see Fig. ... below.

%\subsection{\label{H2O}H2O molecules in a cavity}

The energetics of a rovibrating H$_2$O molecule interacting with an IR cavity mode has been investigated in detail in Ref. \cite{23Sz}. Here we just recall that similar to the case of HCl above, for moderate light-matter coupling polariton formation occurs in the vicinity of well-defined cavity parameters, when polarization induced energy shifts align the energy levels to be nearly resonant with the cavity mode.
In addition, including the rotational fine structure into the molecular model revealed that due to the optical selection rules, only a portion of the near degenerate energy levels of the ``H$_2$O + IR cavity mode" system can interact and form polaritons \cite{23Sz}.
Here we inspect the energetics when adding additional H$_2$O molecules to the cavity, and find conclusions similar to those found in the case of HCl molecules.

\begin{figure}[!ht]
\centering \includegraphics[width=0.35\textwidth]{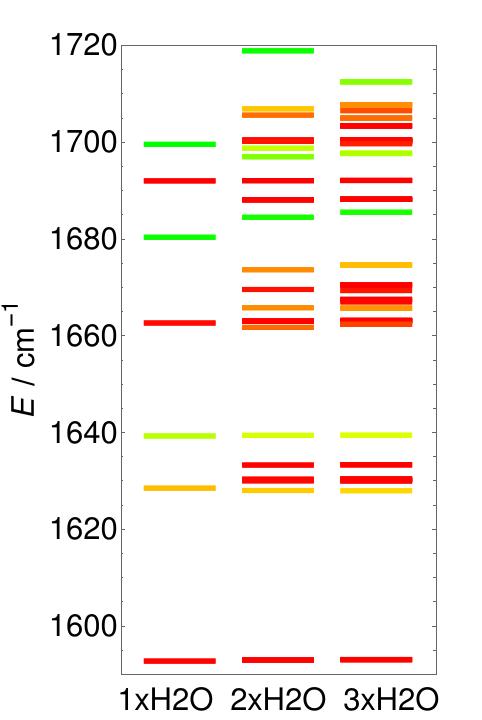}
\centering \includegraphics[width=0.35\textwidth]{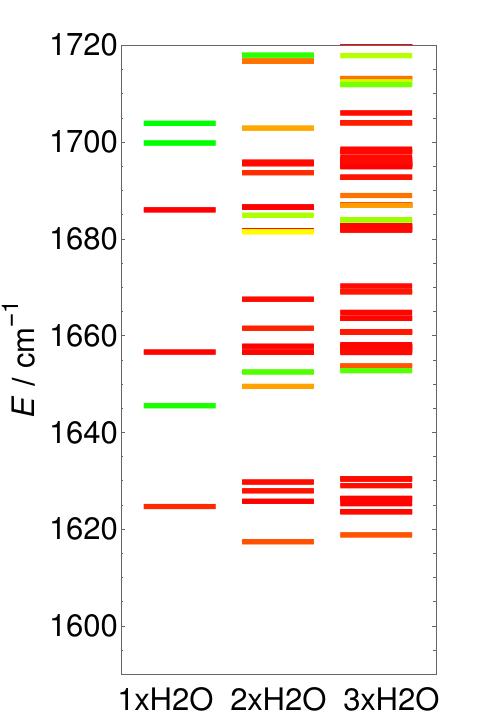}
\caption{Energy levels of the ``$n\times$H$_2$O + IR cavity mode" system for $\hbar \omega_c = 1630$ cm$^{-1}$. Left panel: $g = 200/\sqrt{N_{\rm mol}}$ cm$^{-1}$, right panel: $g = 400/\sqrt{N_{\rm mol}}$ cm$^{-1}$. The colors of the lines represent their character: red indicates zero expectation value for the photon number, while green represents one photon expectation value. Yellow indicates a mixture of photonic and material excitations.
}
\label{fig:3xH2O_1630}
\end{figure}

The left panel of Fig. \ref{fig:3xH2O_1630} shows that at the moderate coupling strengh of $g=200$ cm$^{-1}$ the primary effect of increasing the molecule number is the increase in the available energy levels, nonetheless, energy shift reflecting collective effects are also visible, for example the orange energy level near 1625 cm$^{-1}$ exhibits a slight, barely visible red shift, while the green energy level near 1680 cm$^{-1}$ is considerably shifted.
At the larger coupling strength of $g=400$ cm$^{-1}$ the majority of energy levels seem to be significantly shifted with the addition of molecules, indicating large collective effects.

%\subsection{\label{spectra}High-resolution absorbtion spectra}

\subsection{\label{HCl_spectra}Absorbtion spectra of HCl molecules in a cavity}

Now we look at the absorbtion spectrum of the ``HCl + cavity mode" and ``2$\times$HCl + cavity mode" systems near the vibrational fundamental of HCl when $\hbar \omega_{\rm c}=2925$ cm$^{-1}$ and $g = 650$ cm$^{1}$.
For clarity we assume $T=0$ K spectra, because if the initial state is not the absolute ground state, then deciphering the spectra can become increasingly difficult and counterproductive for identifying the key physical processes responsible for the spectral features.
For 1$\times$HCl the spectra with $\ket{0 J}$ excited initial states might not become too complicated, but for 2$\times$HCl we already have for the first excited state $\approx \ket{0}\red{\ket{0 1}}\blue{\ket{0 0}} \pm \ket{0}\red{\ket{0 0}}\blue{\ket{0 1}}$, which gives a superposition spectrum from the $\ket{0 0}$ and $\ket{0 1}$ molecular components.

\begin{figure}[!ht]
\centering \includegraphics[width=0.45\textwidth]{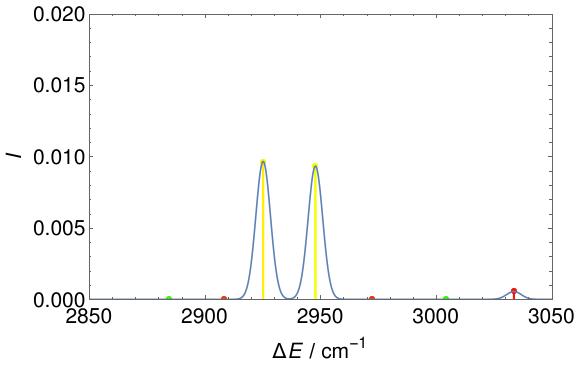}
\centering \includegraphics[width=0.45\textwidth]{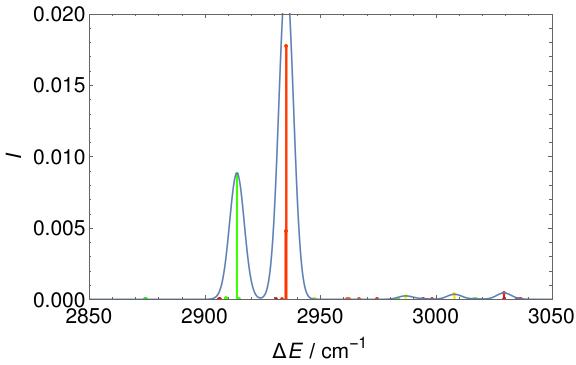}
\centering \includegraphics[width=0.45\textwidth]{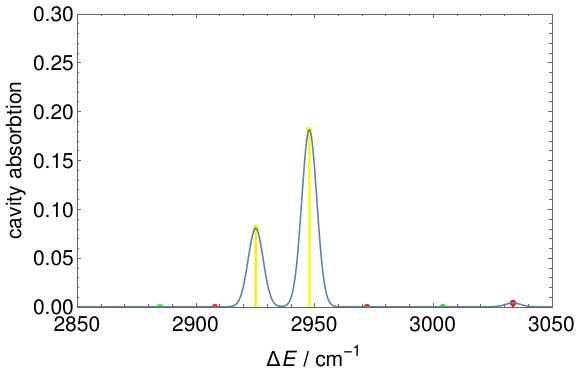}
\centering \includegraphics[width=0.45\textwidth]{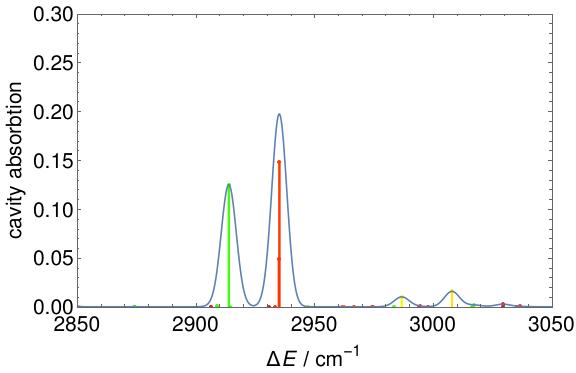}
\caption{Molecular (upper row) and cavity (lower row) absorbtion spectra of the ``1$\times$HCl + cavity mode" (left column) and ``2$\times$HCl + cavity mode" (right column) systems. The initial states in the depicted transitions are primarily composed of $\ket{0}\ket{0 0}$ (left column) and $\ket{0}\red{\ket{0 0}}\blue{\ket{0 0}}$ (right column). The coupling strengh and cavity photon energy are $g=650/\sqrt{N_{\rm mol}}$ cm$^{-1}$ and $\hbar \omega_c = 2925$ cm$^{-1}$, respectively.}
\label{fig:nxHCl_in_cavity_spectra}
\end{figure}

\begin{figure}[!ht]
\centering \includegraphics[width=0.5\textwidth]{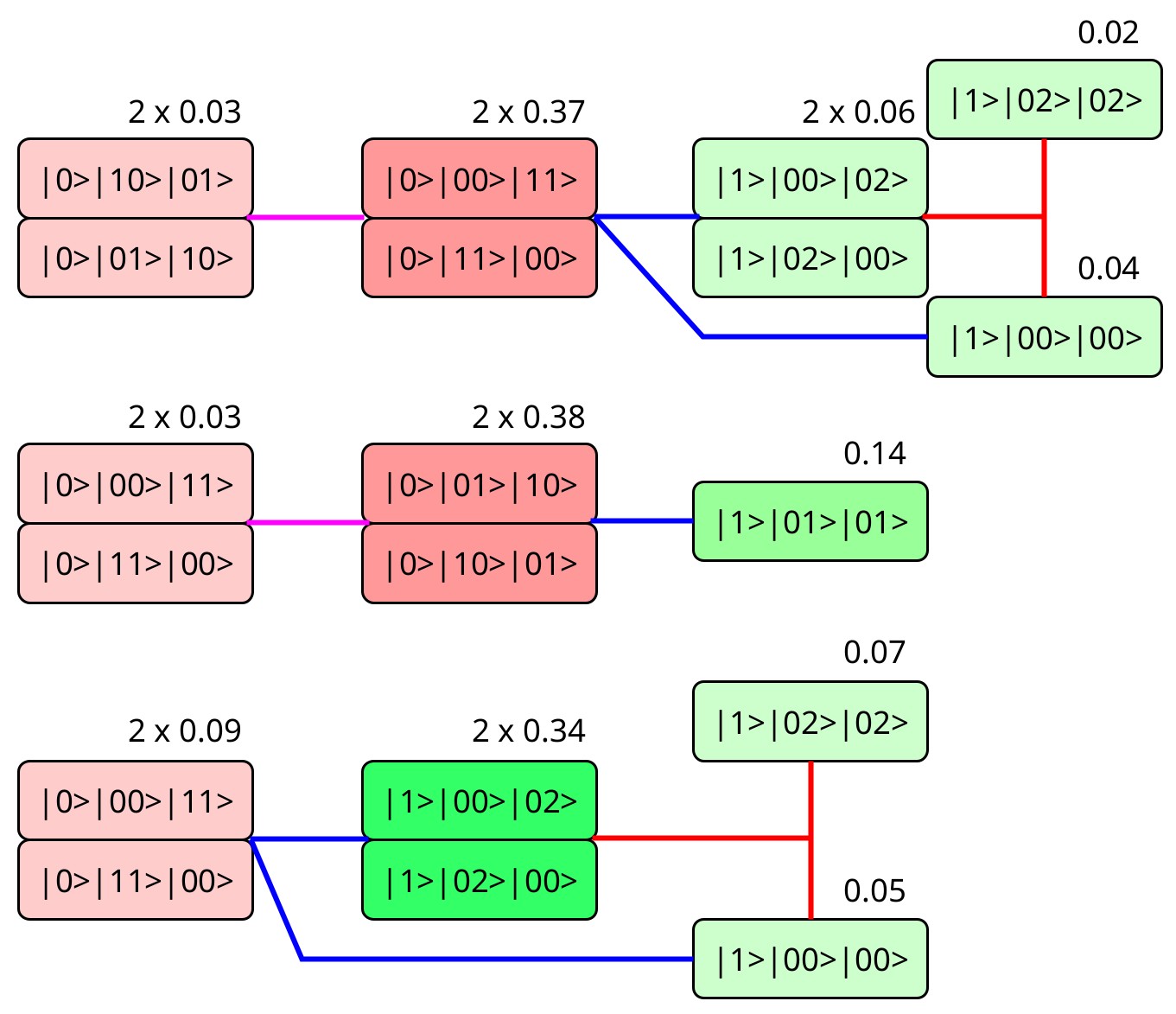}
\caption{Composition of the collective polaritonic states of the ``$2\times$HCl + IR cavity mode" system located at 2914.0 cm$^{-1}$ (lower graph), 2935.1 cm$^{-1}$ (middle graph), and 2935.3 cm$^{-1}$ (upper graph). The weight of the basis functions are displayed above the boxes, and the blue, red, and purple lines connecting the boxes represent the dipole, polarizability, and self-dipole interactions between the different basis functions, respectively. In order to also have a simple visual representation, the boxes are colored according to whether the basis function contains photonic (green) or vibrational (red) excitation, the intensity of the colors represent the weight of the basis function in the polaritonic state. The coupling strengh and cavity photon energy are $g=650/\sqrt{2}$ cm$^{-1}$ and $\hbar \omega_c = 2925$ cm$^{-1}$, respectively.}
\label{fig:2xHCl_three_polariton_SN}
\end{figure}

Fig. \ref{fig:nxHCl_in_cavity_spectra} shows the computed molecular and cavity absorbtion spectra.
Comparing the two columns of Fig. \ref{fig:nxHCl_in_cavity_spectra} reveals major differences between the ``1$\times$HCl and ``2$\times$HCl" spectra.
For two HCl molecules the self-dipole interaction introduces intermolecular couplings where for both molecules $\Delta J=1$ with $\Delta N=0$ for the cavity.
With the large $g$ values used, these additional couplings can lead to noteworthy effects in the collective states, and although the major transitions still can be identified as $\ket{1 1} \leftarrow \ket{0 0}$, the number and position of lines differ.
For example, in the upper right panel of Fig. \ref{fig:nxHCl_in_cavity_spectra} one can identify three absorbtion lines, with the two higher-lying lines almost degenerate.
The final collective states of these three transitions are inspected in Fig. \ref{fig:2xHCl_three_polariton_SN}, showing the basis functions contributing the most to the collective states, as well as the major interactions coupling the basis functions.
Fig. \ref{fig:2xHCl_three_polariton_SN} reveals the appearance of the intermolecular interaction introduced by the self-dipole term, giving rise to the middle (bright) polariton at 2935.1 cm$^{-1}$.
The primary components in this polariton are $\ket{0}\red{\ket{01}}\blue{\ket{10}}$, $\ket{0}\red{\ket{10}}\blue{\ket{01}}$ and $\ket{1}\red{\ket{01}}\blue{\ket{01}}$ with weights 0.38, 0.38 and 0.14, respectively.
These states are optically forbidden from the $\ket{0}\red{\ket{00}}\blue{\ket{00}}$ initial state, however, the self-dipole interaction mixes the optically allowed $\ket{0}\red{\ket{11}}\blue{\ket{00}}$ and $\ket{0}\red{\ket{00}}\blue{\ket{11}}$ components into the polariton, with 0.03 weights, i.e., the self-dipole term makes the polariton bright through an intensity borrowing effect \cite{Cederbaum_multimode,vibronic_coupling_model_Cederbaum_AnnRevPhysChem_2004,CI_spectroscopyYarkony_AnnRevPhysChem_2012}.
The fingerprint of the self-dipole mediated intermolecular interaction in the high-resolution spectra resonantes with recent work \cite{Sidler2024}, where such interaction was shown to be responsible for large local changes in chemical reactivity inside an IR cavity.

The lower polariton at 2914.0 cm$^{-1}$ is primarily composed of $\ket{1}\red{\ket{02}}\blue{\ket{00}}$ and $\ket{1}\red{\ket{00}}\blue{\ket{02}}$ which are dipole-connected to the optically allowed $\ket{0}\red{\ket{11}}\blue{\ket{00}}$ and $\ket{0}\red{\ket{00}}\blue{\ket{11}}$ states, and are also connected to $\ket{1}\red{\ket{02}}\blue{\ket{02}}$ and $\ket{1}\red{\ket{00}}\blue{\ket{00}}$ through the polarizability interaction.
The composition of the upper polariton at 2935.3 cm$^{-1}$ is similar to that of the lower polariton, but in this case $\ket{0}\red{\ket{11}}\blue{\ket{00}}$ and $\ket{0}\red{\ket{00}}\blue{\ket{11}}$ have the largest weights, and the states $\ket{0}\red{\ket{01}}\blue{\ket{10}}$ and $\ket{0}\red{\ket{10}}\blue{\ket{01}}$ are also mixed into the polariton through the self-dipole coupling.
It is worth noting that as can be predicted from Fig. \ref{fig:HCl_gfugges}, at $g = 650$ cm$^{-1}$, due to the energy level shifts, the spectral transitions from the initial state $\ket{0}\ket{0 0}$ show the polariton formed primarily between $\ket{1}\ket{0 2}$ and $\ket{0}\ket{1 1}$, despite $\hbar \omega_{\rm c} = 2925$ cm$^{-1}$ being resonant with the field-free $\ket{11} \leftarrow \ket{00}$ transition.

\subsection{\label{H2O_spectra}Absorbtion spectra of H$_2$O molecules in a cavity}

\begin{figure}[!ht]
\centering \includegraphics[width=0.45\textwidth]{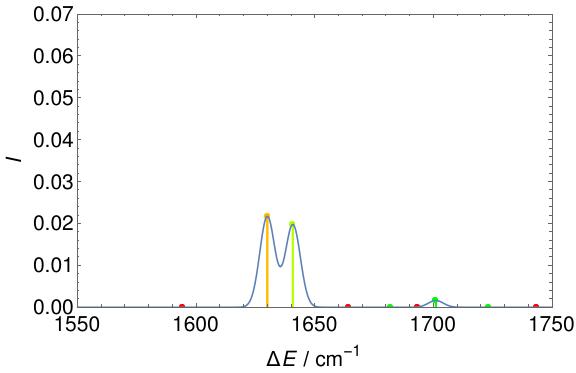}
\centering \includegraphics[width=0.45\textwidth]{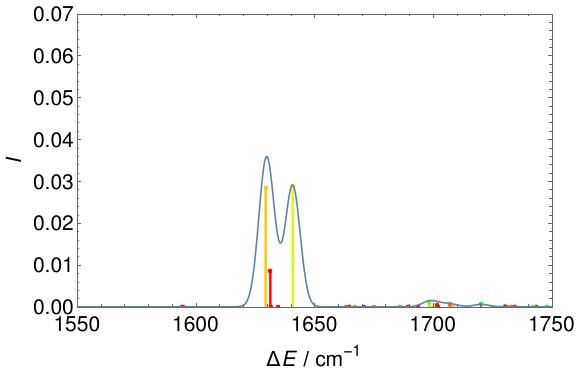}
\centering \includegraphics[width=0.45\textwidth]{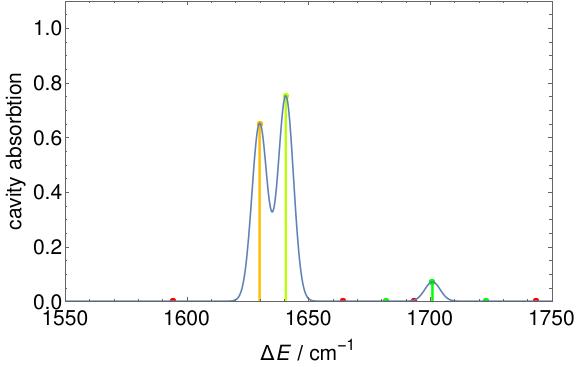}
\centering \includegraphics[width=0.45\textwidth]{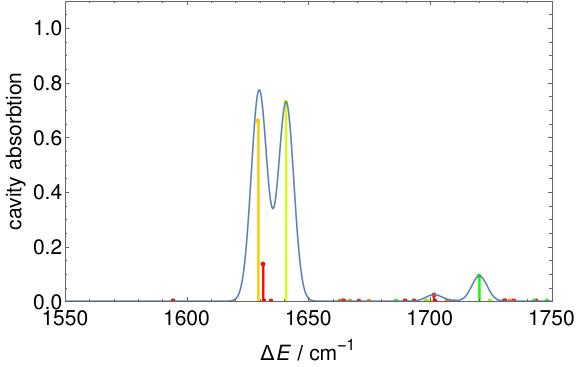}
\caption{Molecular (upper row) and cavity (lower row) absorbtion spectra of the ``1$\times$H$_2$O + IR cavity mode" (left column) and ``2$\times$H$_2$O + IR cavity mode" (right column) systems. The initial states in the depicted transitions are primarily composed of $\ket{0}\ket{(000)(000)}$ (left column) and $\ket{0}\red{\ket{(000)(000)}}\blue{\ket{(000)(000)}}$ (right column). The coupling strengh and cavity photon energy are $g=200/\sqrt{N_{\rm mol}}$ cm$^{-1}$ and $\hbar \omega_c = 1630$ cm$^{-1}$, respectively.}
\label{fig:nxH2O_in_cavity_spectra}
\end{figure}

\begin{figure}[!ht]
\centering \includegraphics[width=0.95\textwidth]{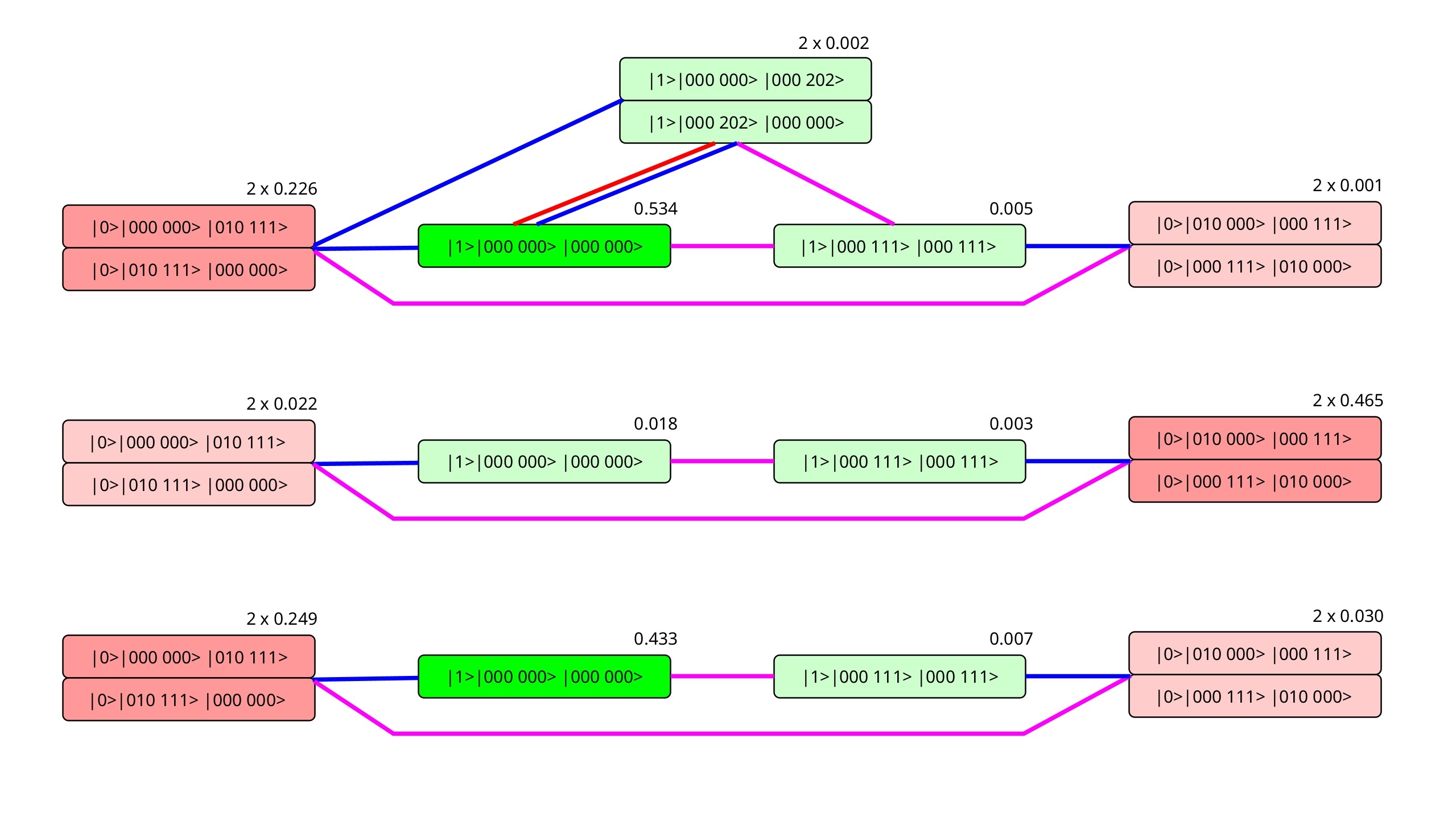}
\caption{Composition of the collective polaritonic states of the ``$2\times$H$_2$O + IR cavity mode" system located at 1629.5 cm$^{-1}$ (lower graph), 1631.5 cm$^{-1}$ (middle graph), and 1640.9 cm$^{-1}$ (upper graph). The weight of the basis functions are displayed above the boxes, and the blue, red, and purple lines connecting the boxes represent the dipole, polarizability, and self-dipole interactions between the different basis functions, respectively. In order to also have a simple visual representation, the boxes are colored according to whether the basis function contains photonic (green) or vibrational (red) excitation, the intensity of the colors represent the weight of the basis function in the polaritonic state. The coupling strengh and cavity photon energy are $g=200/\sqrt{2}$ cm$^{-1}$ and $\hbar \omega_c = 1630$ cm$^{-1}$, respectively.}
\label{fig:2xH2O_three_polariton_SN_g200}
\end{figure}

Fig. \ref{fig:nxH2O_in_cavity_spectra} shows the molecular and cavity absorbtion spectra near the bending fundamental of H$_2$O molecules interacting with a $\hbar \omega_c = 1630$ cm$^{-1}$ cavity mode nearly resonant with the transition $\ket{(010)(111)} \leftarrow \ket{(000)(000)}$ for $g = 200$ cm$^{-1}$ coupling strength.
As for HCl above, the initial state in the computed spectra is the respective ground state of the ``molecule(s)+ cavity mode" system.
%Fig. \ref{fig:nxH2O_in_cavity_spectra} leads to conclusions similar to that found for HCl.
For one H$_2$O molecule, both the molecular and the cavity absorbtion in the left column of Fig. \ref{fig:nxH2O_in_cavity_spectra} shows two prominent peaks reflecting transitions to the two polaritons at 1629.5 cm$^{-1}$ and at 1640.9 cm$^{-1}$, both formed primarily by $\ket{1}\ket{(000)(000)}$ and $\ket{0}\ket{(010)(111)}$.
Adding another H$_2$O molecule, shown in the right column of Fig. \ref{fig:nxH2O_in_cavity_spectra}, leads to no visible change in the peak positions, but modifies the relative peak intensities and gives rise to a bright middle polariton at 1631.5 cm$^{-1}$.
The basis functions having the largest coefficients in the three polaritons of the ``2$\times$H$_2$O + cavity mode" system and the primary interactions between them are shown in Fig. \ref{fig:2xH2O_three_polariton_SN_g200}.
For the upper and lower polariton it seems that at $g = 200$ cm$^{-1}$ coupling strength the self-dipole interaction can only mildly contaminate the most dominant $\ket{0}\red{\ket{(010)(111)}}\blue{\ket{(000)(000)}}$, $\ket{0}\red{\ket{(000)(000)}}\blue{\ket{(010)(111)}}$ and $\ket{1}\red{\ket{(000)(000)}}\blue{\ket{(000)(000)}}$ states with the $\ket{0}\red{\ket{(010)(000)}}\blue{\ket{(000)(111)}}$, $\ket{0}\red{\ket{(000)(111)}}\blue{\ket{(010)(000)}}$ and $\ket{1}\red{\ket{(000)(111)}}\blue{\ket{(000)(111)}}$ states.
This explains the insignificant shift in energy levels, nontheless, even small contaminations can lead to visible changes in the line intensities through intensity borrowing \cite{Cederbaum_multimode,vibronic_coupling_model_Cederbaum_AnnRevPhysChem_2004,CI_spectroscopyYarkony_AnnRevPhysChem_2012}, which is of course most clearly seen in the middle polariton, where the optically forbidden $\ket{0}\red{\ket{(010)(000)}}\blue{\ket{(000)(111)}}$ and $\ket{0}\red{\ket{(000)(111)}}\blue{\ket{(010)(000)}}$ states become visible through their contamination with $\ket{0}\red{\ket{(010)(111)}}\blue{\ket{(000)(000)}}$ and $\ket{0}\red{\ket{(000)(000)}}\blue{\ket{(010)(111)}}$.

%At larger, $g = 400$ cm$^{-1}$ coupling strengh we see that ... \red{SN és spektrum ábrák kellenek ehhez}

%\subsection{\label{dynamics}Energy transfer}
\subsection{\label{HCl_dynamics}Energy transfer in ``HCl molecules + cavity mode" systems}
\begin{figure}[!ht]
\centering \includegraphics[width=0.75\textwidth]{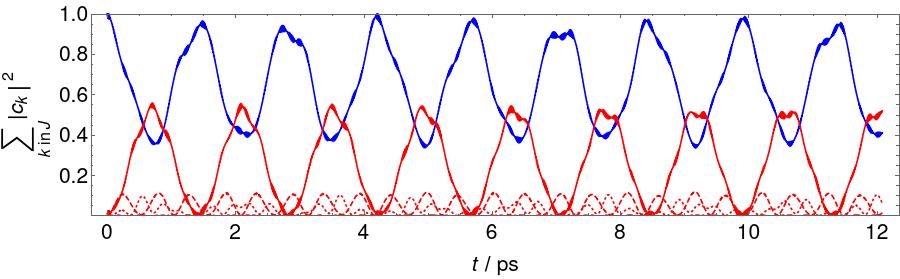}
\centering \includegraphics[width=0.75\textwidth]{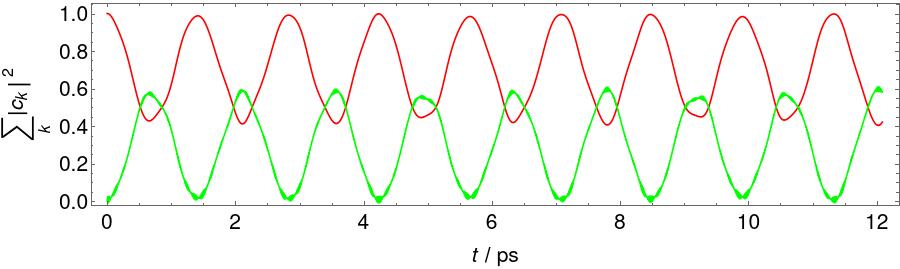}
\centering \includegraphics[width=0.75\textwidth]{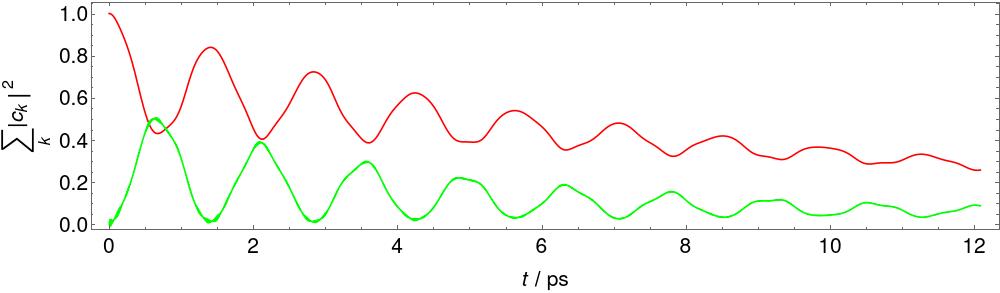}
\centering \includegraphics[width=0.75\textwidth]{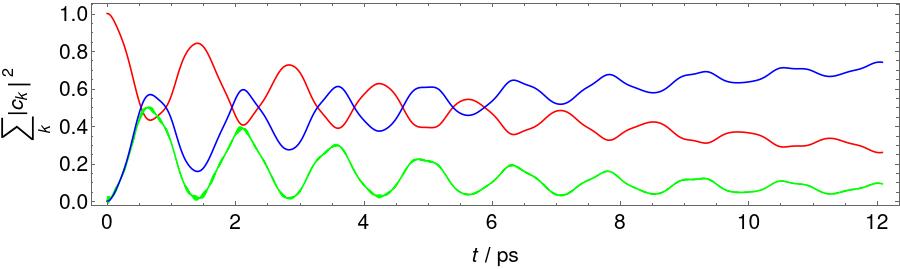}
\caption{Quantum dynamics of the ``1$\times$HCl + cavity" system, with $g=650$ cm$^{-1}$, $\hbar \omega_c = 2925$ cm$^{-1}$, starting from the initial state $\ket{0}\ket{10}$. The depicted curves show sums of basis function populations, according to the following. First row: $J=0$ (blue), $J=1$ (red continous), $J=2$ (red dashed), $J=3$ (red dotted), and $J=4$ (red dotdashed) curves. Second, third and fourth row: vibrationally excited (red curve), photonic excitation (green), vibrational and photonic ground state (blue). First two rows: Hermitian dynamics, third and fourth row: non-Hermitian and Lindblad dynamics, respectively, with $\gamma_c = 2$ cm$^{-1}$ cavity leakage.}
\label{fig:1xHCl_excdyn_init10_g650}
\end{figure}

\begin{figure}[!ht]
\centering \includegraphics[width=0.75\textwidth]{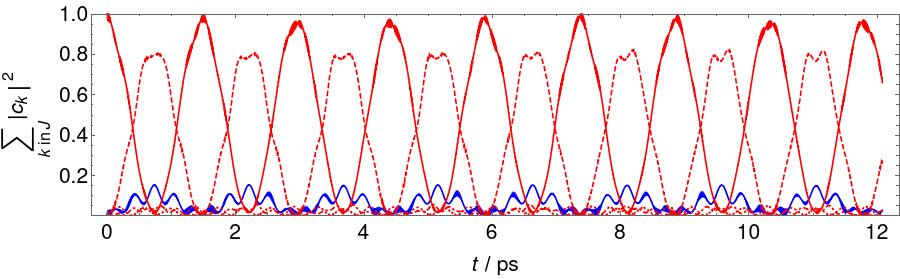}
\centering \includegraphics[width=0.75\textwidth]{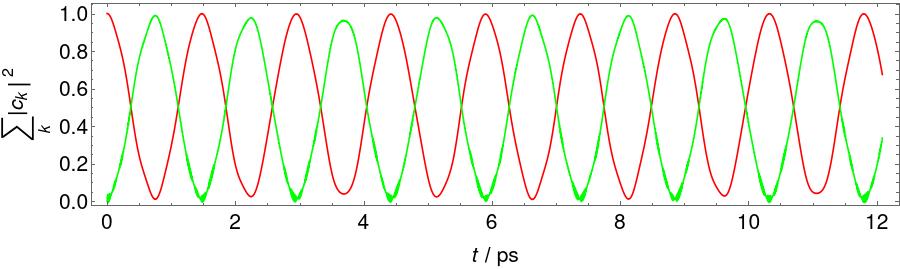}
\centering \includegraphics[width=0.75\textwidth]{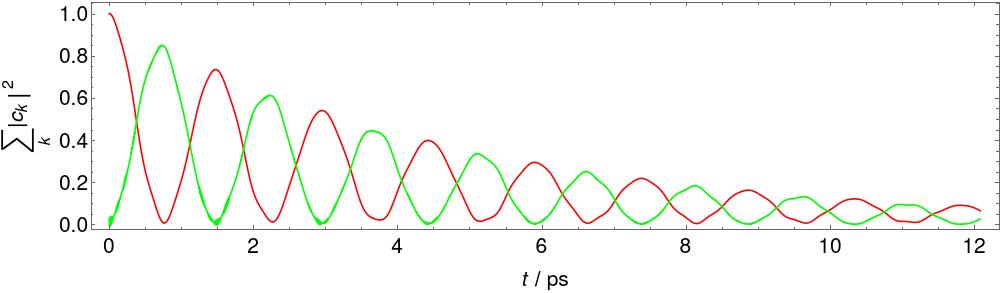}
\centering \includegraphics[width=0.75\textwidth]{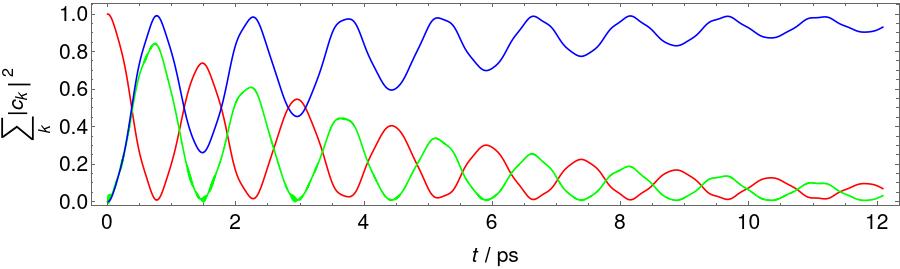}
\caption{Same as Fig. \ref{fig:1xHCl_excdyn_init10_g650}, but with the initial state $\ket{0}\ket{11}$.}
\label{fig:1xHCl_excdyn_init11_g650}
\end{figure}

Let us now turn to the time-dependent quantum dynamics of our systems and investigate the energy transfer between the subsystems and their various degrees of freedom.
Once again we start with the simplest case of one HCl interacting with an $\hbar \omega_c = 2925$ cm$^{-1}$ IR cavity mode nearly resonant with the $\ket{11} \leftarrow \ket{10}$ transition.
Figs. \ref{fig:1xHCl_excdyn_init10_g650} and \ref{fig:1xHCl_excdyn_init11_g650} summarize the simulation results, when the initial state of the system is $\ket{0}\ket{10}$ and $\ket{0}\ket{11}$, respectively, i.e., in both cases initially there are no photons in the field and the HCl molecule is (ro)vibrationally excited.
The second panels from the top in Figs. \ref{fig:1xHCl_excdyn_init10_g650} and \ref{fig:1xHCl_excdyn_init11_g650} show the Hermitian quantum dynamics of the vibrational and photonic excitations, with the curves obtained as the sum of all populations from states having molecular vibrational of photonic excitation, respectively.
In both cases we see oscillatory energy exchange between the two degrees of freedom with a period of around 1.75 ps, and the amplitude of energy exchange is larger for the $\ket{0}\ket{11}$ initial state.
The uppermost panels in Figs. \ref{fig:1xHCl_excdyn_init10_g650} and \ref{fig:1xHCl_excdyn_init11_g650} show the temporal evolution of the sum of populations in the different rotationally excited states.
These plots demonstrate that the efficient energy transfer in these simulations primarily occurs with transitions involving two rotational quantum numbers ($J=0$ and $J=1$ for the $\ket{0}\ket{10}$ initial state and $J=1$ and $J=2$ for the $\ket{0}\ket{11}$ initial state), nonetheless, states with other $J$ values contribute as well.
In their third and fourth rows, Figs. \ref{fig:1xHCl_excdyn_init10_g650} and \ref{fig:1xHCl_excdyn_init11_g650} also show the temporal evolution of the vibrational and photonic excitations when a $\gamma_c=2$ cm$^{-1}$ cavity dissipation is included, either using a non-Hermitian Hamiltonian or solving the Lindblad equations.
For the rovibrational transitions and corresponding resonant photon energies near 3000 cm$^{-1}$, the chosen $\gamma_c$ is in the weak coupling regime.
Based on the $e^{- \gamma_c t}$ formula for the probability of having one photon in the cavity, the cavity lifetime is $1 / \gamma_c \approx 109737$ in atomic units, which is approximately 2.65 ps.
Note that in all simulations shown in this work the lifetime of the excited cavity mode is significantly enhanced through its interaction with the molecule(s).
In agreement with previous results on electronic polaritons \cite{24FaCsaHaCe}, the non-Hermitian Hamiltonian approach works perfectly for the excited state dynamics as long as no population is generated in the manifold where the dynamics takes place, i.e., in the absence of decay from higher lying excited states and no pumping from the ground state manifold.

\begin{figure}[!ht]
\centering \includegraphics[width=0.75\textwidth]{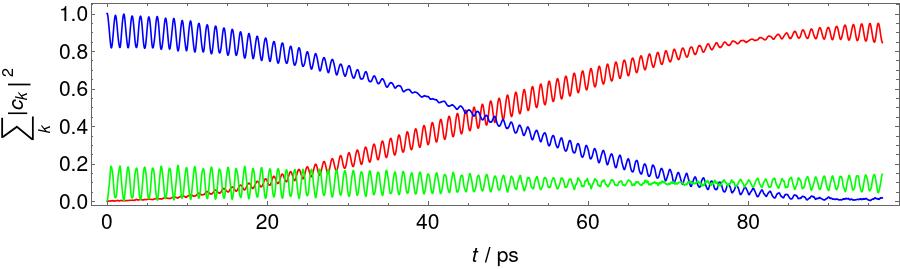}
\centering \includegraphics[width=0.75\textwidth]{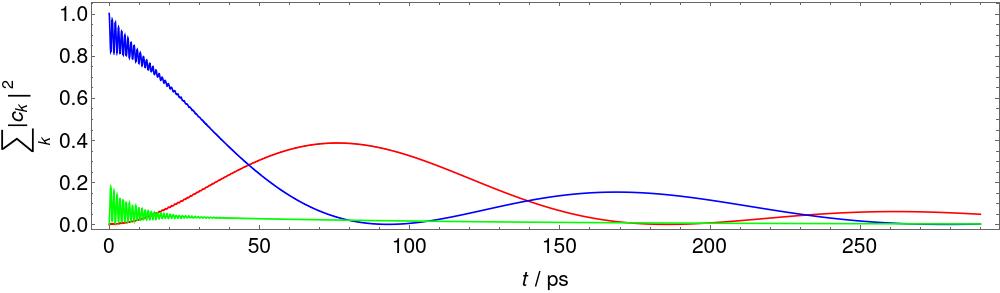}
\caption{Quantum dynamics of the ``2$\times$HCl + cavity" system, with $g=650/\sqrt{2}$ cm$^{-1}$, $\hbar \omega_c = 2925$ cm$^{-1}$, starting from the initial state $\ket{0}\red{\ket{00}}\blue{\ket{10}}$. The depicted curves show sums of basis function populations: molecule one is vibrationally excited (red curve), photonic excitation (green), molecule two is vibrationally excited (blue). First row: Hermitian dynamics, second row: non-Hermitian with $\gamma_c = 2$ cm$^{-1}$ cavity leakage.
}
\label{fig:2xHCl_excdyn_init10_g650}
\end{figure}

\begin{figure}[!ht]
\centering \includegraphics[width=0.75\textwidth]{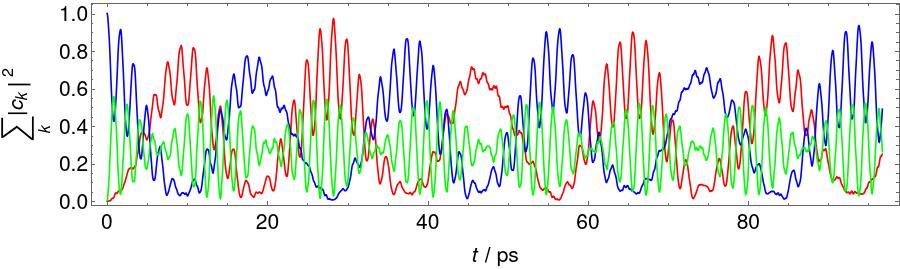}
\centering \includegraphics[width=0.75\textwidth]{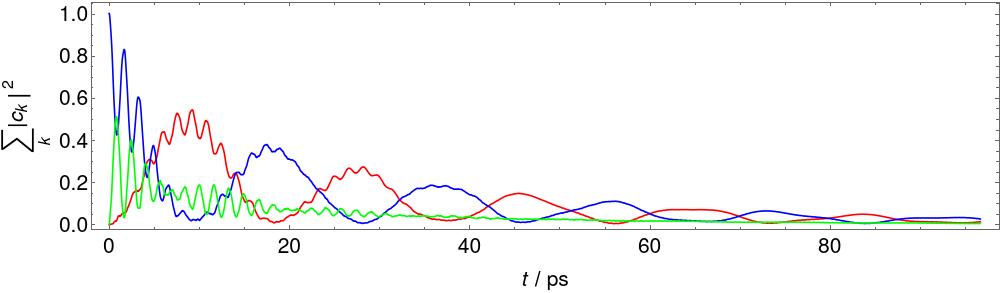}
\caption{Same as Fig. \ref{fig:2xHCl_excdyn_init10_g650} but with the initial state $\ket{0}\red{\ket{00}}\blue{\ket{11}}$.}
\label{fig:2xHCl_excdyn_init11_g650}
\end{figure}

Now we turn to the excitation dynamics when two HCl molecules simultaneously interact with the cavity mode, for which the results are summarized in Figures \ref{fig:2xHCl_excdyn_init10_g650} and \ref{fig:2xHCl_excdyn_init11_g650}, showing population dynamics for the $\ket{0}\red{\ket{00}}\blue{\ket{10}}$ and $\ket{0}\red{\ket{00}}\blue{\ket{11}}$ initial states, respectively.
The upper panels of Figs. \ref{fig:2xHCl_excdyn_init10_g650} and \ref{fig:2xHCl_excdyn_init11_g650} reveal that being confined in an appropriate cavity indeed promotes energy transfer among the two HCl molecules, i.e., with the help of the photonic mode, vibrational excitation is transfered between the two molecules.
This energy transfer can be seen to occur on different timescales, due to various different transition pathways involving different interaction types and rotational levels.
When cavity leakage is included in the model, as shown in the lower panels of Figs. \ref{fig:2xHCl_excdyn_init10_g650} and \ref{fig:2xHCl_excdyn_init11_g650}, the fast, ps timescale oscillations decay in around 20 ps, while slower, 10-100 ps oscillations remain much longer.
This indicates that the fast oscillations originate from transitions involving photonic excitations, while the slow energy exchange in principle could occur through the self-dipole or the $\Delta N = 0$, $\Delta J = 0,\pm2$ polarizability interactions involving states with zero photons in the radiation field.
Since the dynamics seem to be dominated by interactions having $\Delta J = 1$, the self-dipole term is the more likely candidate responsible for the slow oscillations.

\clearpage

\subsection{\label{H2O_dynamics}Energy transfer in ``H$_2$O molecules + cavity mode" systems}

\begin{figure}[!ht]
\includegraphics[width=0.75\textwidth]{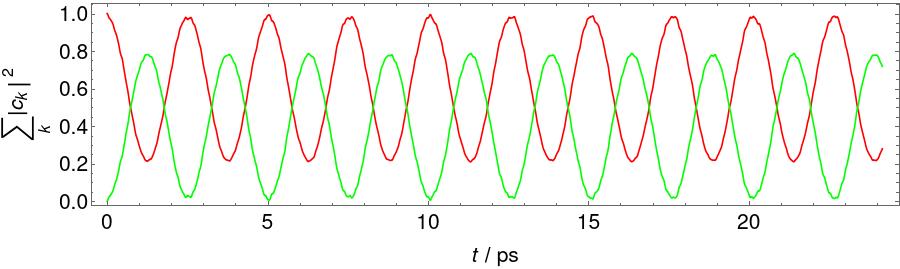}
\includegraphics[width=0.75\textwidth]{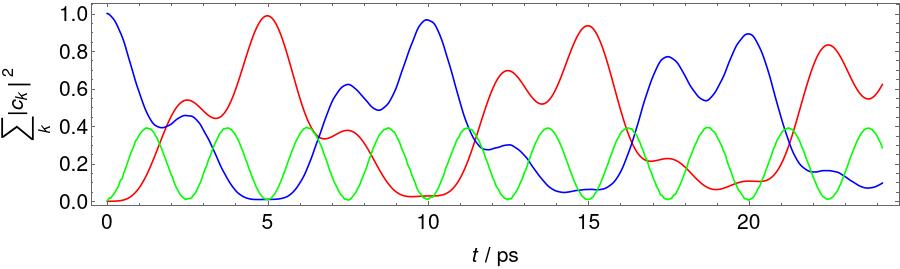}
\includegraphics[width=0.75\textwidth]{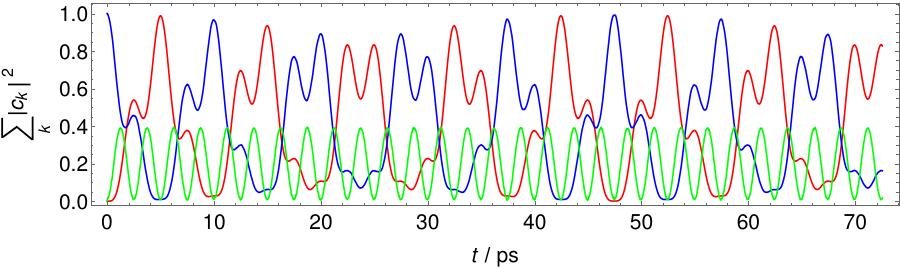}
\caption{Hermitian quantum dynamics of the ``$N_{\rm mol}\times$H$_2$O + IR cavity" system, with $g=200/\sqrt{N_{\rm mol}}$ cm$^{-1}$, $\hbar \omega_c = 1630$ cm$^{-1}$, starting from the initial states $\ket{0}\ket{(010)(111)}$ and $\ket{0}\red{\ket{(000)(000)}}\blue{\Ket{(010)(111)}}$ for $N_{\rm mol}=1$ and 2, respectively. The depicted curves show sums of basis function populations; upper panel: molecule is vibrationally excited (red curve) or photonic excitation (green curve), middle and lower panels: molecule one is vibrationally excited (red curve), photonic excitation (green curve), molecule two is vibrationally excited (blue curve). Upper panel: $N_{\rm mol}=1$, middle and lower panels: $N_{\rm mol}=2$.}
\label{fig:nxH2O_excDyn_g200_Herm}
\end{figure}

Figure \ref{fig:nxH2O_excDyn_g200_Herm} shows the temporal evolution of molecular and photonic excitations under unitary Hamiltonian dynamics, when one or two H$_2$O molecules interact with a cavity mode resonant with the $\ket{(010)(111)} \leftarrow \ket{(000)(000)}$ transition.
The initial states are chosen to be $\ket{0}\ket{(010)(111)}$ and $\Ket{0}\red{\Ket{(000)(000)}}\blue{\Ket{(010)(111)}}$.
The upper panel of Figure \ref{fig:nxH2O_excDyn_g200_Herm} demonstrates that efficient energy exchange occurs between the single H$_2$O molecule and the radiation mode, and a full cycle of energy transfer is realized in around 2.5 picoseconds.
This is considerably shorter than the 4.3 ps one gets based on the $\omega_R = \sqrt{\Delta^2 + W^2}/\hbar$ formula for the Rabi frequency, where $\Delta=E_{(010)(111)}-E_{(000)(000)}-\hbar \omega_{\rm c}$ and $ W = \frac{g}{ea_{0}} \Bra{(010)(111)} \hat{\mu}_z \Ket{(000)(000)}$ and $g=200$ cm$^{-1}$.
Our numerical testing showed that the Rabi formula overestimates the period of energy exchange for other $g$ values as well.
The discrepancy can be rationalized considering that H$_2$O is a not a two level system, and the rotational fine structure allows for several transition pathways between the manifolds of the vibrationally excited molecule and the excited radiation field.
The middle and bottom panels of Fig. \ref{fig:nxH2O_excDyn_g200_Herm} show an identical timescale for the oscillations in the photonic excitation, but efficient intermolecular vibrational energy transfer between the two H$_2$O molecules also appears, and the additional transition pathways lead to multiple beating frequencies.
The full oscillation time of the intermolecular vibrational energy transfer appears to be around 10 ps, which is modulated by the 2.5 ps time period signal and a very slow oscillation having several tens of picoseconds time period.

\begin{figure}[!ht]
\includegraphics[width=0.75\textwidth]{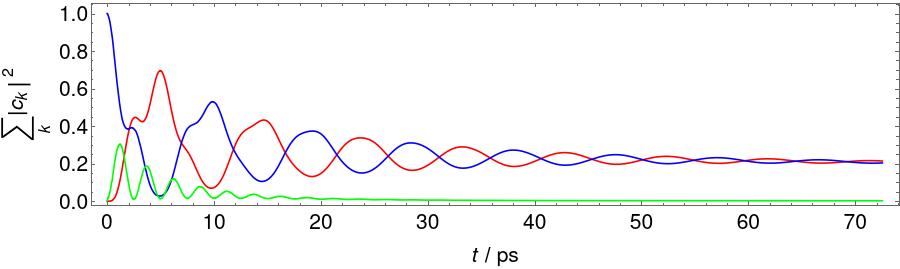}
\includegraphics[width=0.75\textwidth]{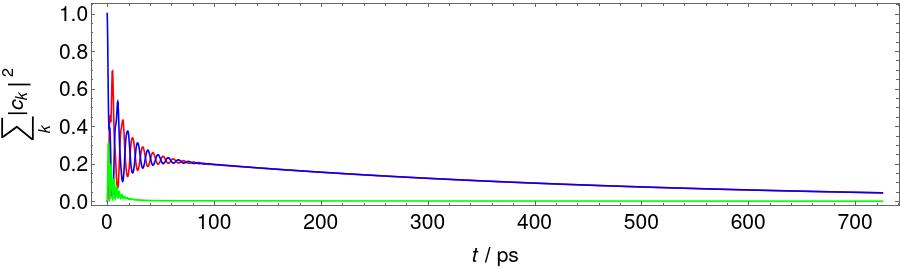}
\caption{Non-Hermitian quantum dynamics of the ``$2\times$H$_2$O + IR cavity" system, with $g=200/\sqrt{2}$ cm$^{-1}$, $\hbar \omega_c = 1630$ cm$^{-1}$, and $\gamma_c = 2$ cm$^{-1}$ cavity leakage, starting from the initial state $\ket{0}\red{\ket{(000)(000)}}\blue{\Ket{(010)(111)}}$. The depicted curves show sums of basis function populations: molecule one is vibrationally excited (red curve), photonic excitation (green curve), molecule two is vibrationally excited (blue curve).}
\label{fig:nxH2O_excDyn_g200_nonHerm}
\end{figure}

Figure \ref{fig:nxH2O_excDyn_g200_nonHerm} shows the temporal evolution of molecular and photonic excitations when cavity leakage is incorporated into the model.
The time period for vibrational energy transfer doesn't seem to be modified by the dissipation, and as shown in the upper panel of Figure \ref{fig:nxH2O_excDyn_g200_nonHerm}, both molecules remain vibrationally excited even after the cavity mode is relaxed completely.
Analyzing the wave function reveals that after several tens of picoseconds of dissipative dynamics the wave function collapses almost completely to the $\vert \psi_d \rangle \propto \Ket{0}\red{\Ket{(010)(111)}}\blue{\Ket{(000)(000)}} - \Ket{0}\red{\Ket{(000)(000)}}\blue{\Ket{(010)(111)}}$ dark state, hindering further energy loss through the cavity mode.
Nonetheless, because the molecules are not two-level systems, the dark state is not completely dark, i.e., dissipation does not vanish completely, but proceeds on a slower time scale, as shown in the bottom panel of Figure \ref{fig:nxH2O_excDyn_g200_nonHerm}.

\subsection{\label{sec:CH4}Connection to experimental results on CH$_4$}

\begin{figure}[!ht]
\includegraphics[width=0.75\textwidth]{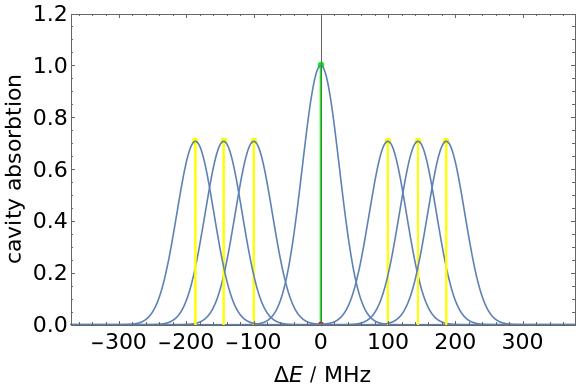}
\caption{Simulated spectra of CH$_4$ in a cavity, reflecting the experimental parameters of Ref. \cite{Wright2023}, using a single molecule model and effective coupling strenghs of 0.154421$\times \sqrt{\rm [CH_4]}/\sqrt{3.5}$ cm$^{-1}$, where ${\rm [CH_4]}=$ 0, 1, 2.1 and 3.5 were used. The computed transition peaks are convolved with a gaussian function having a full-width half maximum of 65 MHz, the cavity line width in the experiment.}
\label{fig:2level_0_1_2p1_3p5_nMol1}
\end{figure}

In this section the theory detailed in Ref. \cite{23Sz} and above is used to simulate the cavity transmission spectra of CH$_4$ using parameters taken directly from the experiments.
Using the data from the supplementary information of Ref. \cite{Wright2023}, the $g=e a_{0} \sqrt{\hbar\omega_{{\rm c}}/(2 \varepsilon_{0}V)}$ single molecule coupling strengh gives 1.80129$\times 10^{-30}$ J.
This computed $g$ value is multiplied with the square root of the number of coupled molecules to obtain the effective coupling strengh used in the simulations.
The effective coupling strengh value is 0.154421 cm$^{-1}$ for the largest CH$_4$ concentration employed in Ref. \cite{Wright2023}.
For such a small coupling strength all $O(g^2)$ and higher order terms can safely be neglected, which means that the collective effects observed for HCl and H$_2$O above will not appear in this case.
Adopting a two-level model of CH$_4$, involving the initial and final states of the $\nu_3$ $J=3 \rightarrow 4$ A$_2$(0) transition, and employing a single cavity mode resonant with this transition was used to compute the cavity spectra shown in Fig. \ref{fig:2level_0_1_2p1_3p5_nMol1}.
Fig. \ref{fig:2level_0_1_2p1_3p5_nMol1} shows that the Rabi splitting scales with molecule number as in the experiment, but theory seems to somewhat underestimate the absolute value of the splitting (compare to Fig. 3(a) of Ref. \cite{Wright2023}).
The origin of this mismatch is not clear, it might originate from incorrectly adopting the experimental parameters to theory, from the limitations of the model itself, i.e., computation of the transitions do not incorporate line broadening, possible error in the computational value of the transition dipole, etc.
Also note that the absorbtion of the uncoupled or dark CH$_4$ gas, leading to the attenuation of the experimental signal near the field-free CH$_4$ absorbtion \cite{24Schwennicke}, is absent from the simulation.

\begin{figure}[!ht]
\includegraphics[width=0.75\textwidth]{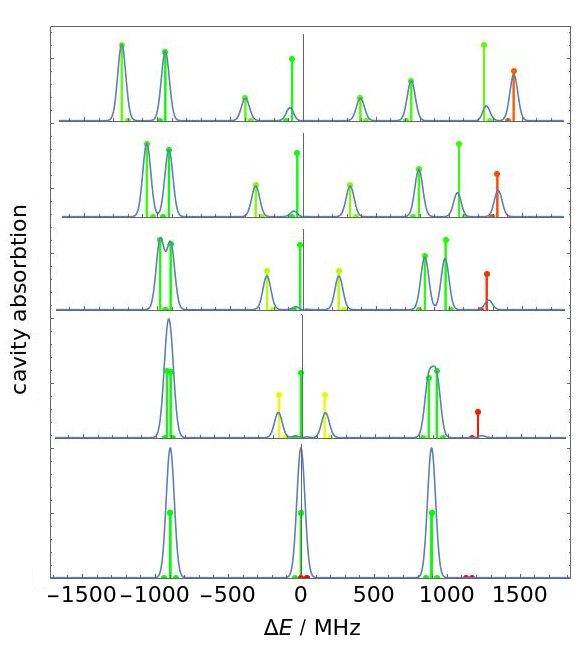}
\caption{Simulated spectra of CH$_4$ in a cavity, reflecting experimental parameters from Ref. \cite{Wright2023_b}, using a single molecule model and effective coupling strenghs of 0.154421$\times \sqrt{\rm [CH_4]}$/$\sqrt{3.5}$ cm$^{-1}$, where ${\rm [CH_4]}=$ 0, 2.6, 7.2, 15 and 30 were used. The computed transition peaks are convolved with a gaussian function having a full-width half maximum of 87 MHz. The absorbtion by uncoupled/dark CH$_4$ is included by decreasing the spectrum envelope at the field-free CH$_4$ transition frequencies with a gaussian having a full-width half maximum of 282 MHz.}
\label{fig:4level_real_spectrum_nMol1}
\end{figure}

Fig. \ref{fig:4level_real_spectrum_nMol1} was generated using a four-level model of CH$_4$, involving the initial and final states of both the A$_2$(0) and F$_2$(0) transitions investigated in Ref. \cite{Wright2023_b} (F$_2$(0) is labeled F$_1$(0) in Ref. \cite{Kefala2024}), and including three adjacent cavity modes with the middle one centered at the A$_2$(0) transition.
Absorbtion of the uncoupled/dark CH$_4$ gas is also included artificially into the spectrum envelope in Fig. \ref{fig:4level_real_spectrum_nMol1}.
Upon comparison with Fig 2(a) of Ref. \cite{Wright2023_b}, one finds that the simulations of this work reproduce the experimental peak progressions quite well, although there are some discrepancies in the peak heights and positions.

Overall, upon using realistic experimental parameters, the theoretical approach of this work and Ref. \cite{23Sz} is capable of capturing the main features of the experiments of Refs. \cite{Wright2023, Wright2023_b}, without any fitting to the experimental data.
However, the match is not perfect, considering additional aspects of the experiments, possible adjustments to the used parameters, or extending the theory are likely needed for better agreement.
Due to the very small effective light-matter coupling strength exhibited in the experiments, the collective effects introduced by the self-dipole and polarizability terms could not be tested here.

\section{Summary}
In summary, this theoretical work presented how accurate rotationally resolved molecular models can be employed to gain insight in high-resolution into the collective effects and intermolecular processes arising when molecules in the gas phase are interacting with a resonant infrared (IR) radiation mode.
The theoretical approach, expanded in detailed for two molecules in a cavity, shed light on the selection rules of the various cavity-mediated interactions between the field-free molecular eigenstates.
To showcase the capabilities of the theory, numerical results were presented for rovibrating HCl and H$_2$O molecules interacting with a resonant cavity mode.
HCl was shown to be more susceptible to forming purely rotational, rather than rovibrational polaritons.
Collective effects induced by the cavity-mediated interaction between the molecules were identified in the energy level shifts of the ``molecules + cavity mode" systems.
The molecular and cavity absorbtion spectra also demonstrated collective effects via the changes in transition energies, peak intensities, as well as via the appearance of new peaks, occurring as a result of a cavity-mediated intensity borrowing effect.
Hermitian and non-Hermitian quantum dynamics simulations showed that the different intermolecular interactions, mediated by the cavity mode, can act as channels of intermolecular energy transfer involving various rotational states.
In the simulations presented, efficient intermolecular vibrational energy transfer was found to occur on a picosecond timescale with contributions from several interaction channels, leading to a complicated temporal evolution of the populations.
In order to connect to existing experimental work, simulations on CH$_4$ interacting with resonant IR cavity mode(s) were also carried out.
These results showed that upon using realistic experimental parameters, the theoretical approach of this work is capable of capturing the main features of the experiments without any fitting to the experimental data.

\clearpage

\section{Acknowledgements}
This research was supported by the NKFIH National Research, Development and Innovation Office (Grant No. FK134291).
The Author is grateful to Csaba F\'abri for using his in house code to carry out computations on the expectation value of the dipole square components of HCl.

\bibliography{main_master}

\end{document}